\documentclass[pre,singlecolumn,noshowpacs,preprintnumbers,amsmath,amssymb,nofootinbib]{revtex4}

\usepackage{mathrsfs}
\usepackage[pdftex]{graphicx}
\usepackage{dcolumn}
\usepackage{bm}
\usepackage{color}
\usepackage{ascmac}

\usepackage{tcolorbox} 
\tcbuselibrary{skins,breakable,theorems}
\newenvironment{box_normal}{
    \begin{tcolorbox}[	colframe = gray!70, colback = gray!3, 
												coltitle=black, fonttitle=\bfseries, breakable = false ]
}{
    \end{tcolorbox}
}

\newcommand{\mcL}{\mathcal{L}}
\newcommand{\mcK}{\mathcal{K}}
\newcommand{\mcD}{\mathcal{D}}

\newcommand{\mcA}{\mathcal{A}}
\newcommand{\tl}{\tilde}
\newcommand{\tilh}{\tilde{h}}
\newcommand{\ra}{\rangle}
\newcommand{\la}{\langle}
\newcommand{\tb}{\textbf}

\newcommand{\eps}{\epsilon}

\newcommand{\hv}{\hat{v}}
\newcommand{\hta}{\hat{a}}
\newcommand{\hW}{\hat{W}}

\newcommand{\hL}{\hat{L}}

\newcommand{\hxi}{\hat{\xi}}
\newcommand{\hlambda}{\hat{\lambda}}

\newcommand{\hz}{\hat{z}}
\newcommand{\hZ}{\hat{Z}}
\newcommand{\hV}{\hat{V}}
\newcommand{\hy}{\hat{y}}
\newcommand{\hY}{\hat{Y}}
\newcommand{\rG}{\mathrm{G}}

\newcommand{\hGamma}{\hat{\Gamma}}

\newcommand{\heta}{\hat{\eta}}

\newcommand{\ve}{\varepsilon}

\newcommand{\hw}{\hat{w}}
\newcommand{\mrG}{\mathrm{G}}
\newcommand{\mrP}{\mathrm{P}}

\newcommand{\htt}{\hat{t}}

\newcommand{\mft}{\mathfrak{t}}

\newcommand{\hx}{\hat{x}}
\newcommand{\hX}{\hat{X}}

\newcommand{\hN}{\hat{N}}
\newcommand{\hxiCP}{\hat{\xi}^{\mathrm{CP}}}

\newcommand{\mrss}{\mathrm{ss}}
\newcommand{\pd}{\partial}
\newcommand*\dd{\mathop{}\!\mathrm{d}}

\newcommand*\dt{\mathop{}\!\mathrm{d}t}
\newcommand*\dy{\mathop{}\!\mathrm{d}y}
\newcommand*\ds{\mathop{}\!\mathrm{d}s}
\newcommand*\df{\mathop{}\!\mathrm{d}f}

\begin{document}
\title{	
				A standard form of master equations for general non-Markovian jump processes: \\ 
				the Laplace-space embedding framework and asymptotic solution
			}
\author{Kiyoshi Kanazawa}
\affiliation{
			Department of Physics, Graduate School of Science, Kyoto University, Kyoto 606-8502, Japan
		}	
\author{Didier Sornette}
\affiliation{
			Institute of Risk Analysis, Prediction, and Management (Risks-X), Southern University of Science and Technology (SUSTech), Shenzhen 518055, China
		}
\date{\today}

\begin{abstract}
We present a standard form of master equations (ME) for general one-dimensional non-Markovian (history-dependent) jump processes, complemented by an asymptotic solution derived from an expanded system-size approach. 
The ME is obtained by developing a general Markovian embedding using a suitable set of auxiliary field variables. 
 This Markovian embedding uses a Laplace-convolution operation applied to the velocity trajectory.
We introduce an asymptotic method tailored for this ME standard, generalising the system-size expansion for these jump processes. Under specific stability conditions tied to a single noise source, upon coarse-graining, the Generalized Langevin Equation (GLE) emerges as a universal approximate model for point processes in the weak-coupling limit. This methodology offers a unified analytical toolset for general non-Markovian processes, reinforcing the universal applicability of the GLE founded in microdynamics and the principles of statistical physics.
\end{abstract}
\pacs{}

\maketitle
\section{Introduction}

Non-Markovian stochastic processes have emerged as a powerful framework across diverse scientific disciplines, including physics~\cite{KuboB}, chemistry~\cite{VanKampen}, econometrics~\cite{JDHamilton}, and financial modeling~\cite{BouchaudTradebook2018}.
\begin{itemize}
	\item Physics: Within statistical physics, particle motion in water is described by the generalized Langevin equation (GLE)~\cite{KuboB}. The GLE represents a quintessential non-Markovian stochastic model, capturing the hydrodynamic memory effect.

	\item Econometrics: The autoregressive integrated moving average (ARIMA) model stands as a recognized discrete-time non-Markovian model 
	describing many stylized structures of financial returns~\cite{JDHamilton}.

	\item Finance: The self-excited Hawkes process~\cite{Hawkes1,Hawkes2,Hawkes3}, a widely-used non-Markovian point process model, finds many applications in finance~\cite{Hawkes2018,Bowsher2007,Filimonov2012,Bacry2015}. Here, ``points'' indicate event occurrences on the time axis.

	\item Other Disciplines: The versatility of the non-Markovian self-exciting Hawkes process also extends to neuroscience~\cite{Gerhard2017}, seismology~\cite{Ogata1988,Ogata1999,Helmstetter2002,Nandan2019}, epidemiology~\cite{Feng2019,Schoenberg19},
	industrial/organizational Psychology and sociology~\cite{Rametal22,Rametal22b}, criminology \cite{Mohlercrime11} and so on.
\end{itemize}
Central to these models is their ability to encapsulate long-memory effects inherent to various systems. This is typified by power law decaying autocorrelation functions (ACFs), transcending the conventional boundaries set by Markovian stochastic processes.

A well-established analytical toolkit has been developed for Markovian stochastic processes ~\cite{GardinerB,RiskenB}, which includes stochastic differential equations (SDEs), master equations (MEs), and their asymptotic solutions. For instance, the theory of standard forms has been instrumental in the systematic classification of both SDEs and MEs. Given that MEs represent linear time-evolution equations for probability density functions and functionals (PDFs), they can be solved within the framework of linear algebra, particularly through methods like the eigenfunction expansion~\cite{GardinerB,RiskenB}.

There are also various asymptotic methods tailored to MEs. Prominent among these are the system-size expansion~\cite{VanKampen,KzBook,KzPRL,KzJStat} and the Wentzel-Kramers-Brillouin (WKB) approximation~\cite{RiskenB,GrahamTel1984}. Notably, the system-size expansion stands as a historic cornerstone in the realm of statistical physics, especially concerning the Langevin equations. This is largely due to its role in extrapolating various Langevin equations from underlying microscopic physical dynamics. Hence, Markovian process theory offers a robust and structured foundation for statistical physics, at least in a formal sense.

In contrast to the structured theories for Markovian processes, those for non-Markovian processes remain more fragmented. A universally accepted master equation (ME) theory for non-Markovian processes is absent. Current MEs pertain specifically to particular non-Markovian SDE classes, such as GLE with exponential memories~\cite{ZwanzigTB,KupfermanJSP2004}, GLE with linear potential~\cite{Siegle2010}, and semi-Markovian point processes~\cite{KlafterB}. Without a standardized form for these MEs, corresponding asymptotic methods for general non-Markovian processes have yet to emerge. Hence, developing a systematic theory for non-Markovian processes remains a long-standing challenge in statistical physics.
 
Our prior studies have offered partial solutions to this challenge, specifically for linear and nonlinear Hawkes processes~\cite{KzDidier2019PRL,KzDidier2019PRR,KzDidier2021PRL,KzDidier2023PRR}. We have generalised the Markovian embedding approach, transforming a non-Markovian process into a Markovian field dynamic. Within this framework, the MEs for the Hawkes processes are conceived as time-evolution equations for the PDFs of auxiliary field variables. We refer to these equations as {\it field master equations}. While we regard this methodology as a potential avenue for generating MEs for a broader range of non-Markovian processes, its scope, for now, remains confined to certain models, notably the nonlinear extensions of the Hawkes point process family.
 
 In this report, we focus on deriving the master equation (ME) for the general class of one-dimensional non-Markovian jump processes, a subset of the broader point process family. Our approach frames the general one-dimensional non-Markovian jump process as a history-dependent jump process. As a versatile model, it can incorporate any form of historical dependency and represents the most comprehensive one-dimensional non-Markovian jump process conceivable by us. To tackle these processes, we develop a general Markovian embedding using a suitable set of auxiliary field variables. 
 This Markovian embedding uses a Laplace-convolution operation applied to the velocity trajectory
 and allows us to derive the corresponding field ME. Given the capability of this ME to handle all forms of 
 one-dimensional non-Markovian properties, we suggests that it constitutes a standard ME form for general one-dimensional non-Markovian jump processes.
Additionally, we introduce an asymptotic method tailored for this ME standard, generalising the system-size expansion for this jump process. Under specific stability conditions tied to a single noise source, the Generalized Langevin Equation (GLE) emerges as a universal approximate model for point processes in the weak-coupling limit. This methodology offers a unified analytical tool for general non-Markovian processes, underpinned by strong statistical physics validating the GLE's universal applicability.
	
	This report is organised as follows. Section~\ref{sec:notation} presents our mathematical notations. Sec.~\ref{sec:review_Markov} 
gives a concise review of the theories of Markovian stochastic processes, of the corresponding standard form of the ME and of the system-size expansion. 
Sec.~\ref{sec:Model} introduces our model and derives the corresponding field ME via the Laplace-convolution Markovian embedding. Section~\ref{sec:system_size_expansion} describes the system-size expansion for the non-Markovian jump processes 
that allows us to asymptotically derive the GLE. Sec.~\ref{sec:NLHawkesPricing} demonstrates another application of our formalism to a financial-pricing model based on the nonlinear Hawkes processes. Implications and future perspectives of our work are discussed in Sec.~\ref{sec:discussion}.  Sec.~\ref{sec:conclusion} concludes. Seven appendices supplement the main text on technical issues.  

\section{Mathematical notation}\label{sec:notation}
	Let us describe our mathematical notation regarding stochastic variables, sets, and functionals. 
	
	\subsection{Notation for stochastic variables}
		Any stochastic variable carries the hat symbol in the form $\hat{A}$ to distinguish it from the real number $A$. The probability density function (PDF) is denoted by $P_t(A):=P(\hat{A}_t=A)$, implying that the probability for $\hat{A}_t\in [A,A+\dd A)$ is given by $P_t(A)\dd A$. The ensemble average of any stochastic variable $\hat{A}$ is written as $\la\hat{A}\ra:=\int AP_t(A)\dd A$. Using this notation, the PDF can be rewritten as $P_t(A)=\la \delta(A-\hat{A}_t)\ra$ with the Dirac $\delta$ function (see Appendix~\ref{sec:app:dirac-functionalderivative}). 

	\subsection{Notation for sets}
		The set of real numbers and the set of positive integers are denoted by $\bm{R}$ and $\bm{N}$. The set of positive real numbers is denoted by $\bm{R}^+:= \{s \>|\> s > 0, s\in \bm{R}\}$. Here $s$ typically represents the wave number, which should be a real positive number, and we introduce the compact notation  
		\begin{equation}
			\{z(s)\}_{s} := \{z(s) \>|\> s \in \bm{R}^+\}~.
		\end{equation}
		Also, $i$ and $j$ typically represents integers, and we also introduce the corresponding compact notation  
		\begin{equation}
			\{a_i\}_i := \{a_i \>| \> i \in \bm{N}\}.
		\end{equation}
	
	\subsection{Notation for functionals}
		If the argument of a map $f$ is a function $\{z(s)\}_s$, $f$ is called a {\it functional}. A functional is indicated by the square brackets $f[\{z(s)\}_s]$. The functional notation $f[\{z(s)\}_s]$ is sometimes abbreviated as $f[z]$ if its meaning is obvious from the context. For a stochastic field variable $\{\hz_t(s)\}_s$, the corresponding PDF is written as $P_t[z]=P_t[\{z(s)\}_s]$, characterising the probability $P_t[z]\mcD z$  that $\{\hz(s)\}_s \in \prod_{s}[z(s),z(s)+\dd z(s))$, where the functional volume element is $\mcD z:= \prod_s \dd z(s)$. The ensemble average of any functional $f[\hz]$ is written as the path-integral representation
		\begin{equation}
			\la f[\hz]\ra := \int f[z]P_t[z]\mcD z. 
			\label{eq:def:ensemble_average_functional}
		\end{equation}
		On the basis of this notation, the PDF is formally rewritten by $P_t[z] = \la \delta[z-\hz_t]\ra$, where the $\delta$ functional is defined by $\delta[z-\hz_t]:= \prod_{s \in \bm{R}^+}\delta(z(s)-\hz_t(s))$. The concept of derivative can be generalised to the functional derivative, which is denoted by $\delta f[z]/\delta z(s)$ (see Appendix~\ref{sec:app:dirac-functionalderivative} for the detail).

\section{Literature review: Markovian stochastic processes}\label{sec:review_Markov}
	
This section offers a concise overview of the foundational theory of Markovian processes, serving as an introduction for readers less acquainted with Markovian processes and statistical physics. Specifically, we touch upon the standard form of the master equation (ME) for these processes. Experts who are solely focused on our primary findings may bypass this section, as the main results are presented in a standalone, comprehensive format.

	\subsection{Markovian stochastic differential equations}
		Let us consider a one-dimensional stochastic process characterised by the trajectory $\{\hv_s\}_{s\leq t}$, where $t$ is the current time. If the statistics of the infinitesimal future state $\hv_{t+\dt}$ is completely characterised only by the current state $\hv_t$, the stochastic dynamics is said to obey a {\it Markovian stochastic process}. However, a more general class of stochastic models can be considered that cannot be characterised only by the current state $\hv_{t}$. For example, a stochastic model can depend on the full history $\{\hv_s\}_{s\leq t}$. Such stochastic dynamics obey a {\it non-Markovian stochastic process}. This subsection reviews the theory of Markovian stochastic processes, in particular their standard forms and the system-size expansion. 

		\subsubsection{Standard form of the white noise}
			White noise is a noise that is independent of its history. It is formally defined as the derivative of the L\'evy process $\hL_t$, such that $\hxi^{\rm W}_t:=\dd \hL_t/\dt$. According to the L\'evy-It\^o decomposition, any white noise can be decomposed as the sum of the white Gaussian noise and of the white Poisson noise (see Appendix~\ref{app:review_whitenoise} for a review), such that 
			\begin{equation}
				\hxi^{\rm W} = m + \sigma \hxi^{\mrG} + \hxi^{\mrP}_{\lambda(y)},
				\label{eq:Levy-Ito_decomposition}
			\end{equation} 
			where $m$ is the constant drift, $\sigma$ is the standard deviation of the white noise, $\hxi^{\mrG}$ is the white noise, and $\hxi^{\mrP}_{\lambda(y)}$ is the white Poisson noise with intensity distribution function $\lambda(y)$ of the jump size $y$.

		\subsubsection{Standard form of stochastic differential equations}
			Any one-dimensional stochastic process can be constructed from white noise by introducing the {\it state-dependence} into the drift term $a$, standard deviation $\sigma$, and the intensity distribution function (IDF) $\lambda(y)$, such that 
			\begin{equation}
				m \to m(\hv), \>\>\> \sigma \to \sigma(\hv), \>\>\> \lambda(y) \to \lambda(y|\hv).
			\end{equation}
			A one-dimensional stochastic Markovian  process $\{\hv_s\}_{s\leq t}$ obeys the state-dependent SDE
			\begin{equation}
				\frac{\dd \hv_t}{\dt} = m(\hv_t) + \sigma(\hv_t)\hxi^{\mrG} + \hxi^{\mrP}_{\lambda(y|\hv_t)},
				\label{eq:review:SDE_standardForm}
			\end{equation}
			with the It\^o interpretation assumed. We refer to this representation as the {\it standard form of one-dimensional Markovian SDEs}. 
			The Markovian property is expressed by the fact that the right-hand side of Eq.~\eqref{eq:review:SDE_standardForm} depends only on the current state $\hv_t$.

		\subsubsection{Standard form of master equations}
			The SDE~\eqref{eq:review:SDE_standardForm} describes the dynamics of stochastic systems for a single path. While the SDE are intuitive tools, they are not easy to handle because of their general nonlinear structure. For analytical calculations, the ME approach provides more systematic methods based on linear algebra. The ME is the equation 
governing the time-evolution of the PDF $P_t(v):=\la \delta(v-\hv_t)\ra$ as follows
			\begin{equation}
				\frac{\pd P_t(v)}{\pd t} = \mcL P_t(v)
				\label{eq:review:ME}
			\end{equation}
			with a linear operator $\mcL$. The ME corresponding to the SDE~\eqref{eq:review:SDE_standardForm} is given by 
			\begin{equation}
				\frac{\pd P_t(v)}{\pd t} = \left[-\frac{\pd}{\pd v}m(v) + \frac{1}{2}\frac{\pd^2}{\pd v^2}\sigma^2(v)\right]P_t(v) + \int_{-\infty}^\infty \dd y\left[\lambda(y|v-y)P_t(v-y)-\lambda(y|v)P_t(v)\right].
				\label{eq:review:ME_standardForm}
			\end{equation}
			This ME is known to covers all possible one-dimensional Markovian stochastic processes~\cite{GardinerB}, and, thus, is called the {\it standard form of the master equation} in this report. The ME is very useful because it is always a linear dynamical equation\footnote{	
				Indeed, the formal solution of the ME~\eqref{eq:review:ME} is given by 
				$$
					P_t(\nu) = \sum_{i} c_i e^{-\mu_i t}\phi_i(\nu), \>\>\> 
					\mcL \phi_i(\nu) = -\mu_i \phi_i(\nu).
				$$
				with initial condition constants $\{c_i\}_i$, where $\mu_i$ and $\phi_i(v)$ are the $i$th eigenvalue and the corresponding eigenfunction, respectively. 
			} of the PDF $P_t(v)$. In other words, a standard approach to a given Markovian process is to consider its ME and solve the corresponding eigenvalue problem with linear algebra techniques.
			
		\subsubsection{Markovian jump process (history-independent Poisson process)}
			\begin{figure}
				\centering
				\includegraphics[width=85mm]{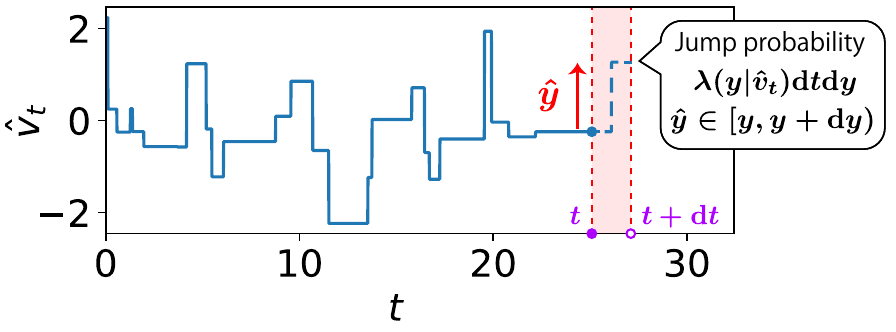}
				\caption{
					Schematic of the Markovian jump process~\eqref{eq:review:Markovian_Jump}. The path is piecewisely continuous and occasionally has jumps. A jump occurs during $[t,t+\dd t)$ with probability $\lambda(y|\hv_t) dt dy$ with jump size $\hy \in [y, y+\dd y)$. The remarkable character of the Markovian jump process is that the intensity density $\lambda(y|\hv_t)$ depends only on the current state $\hv_t$ and does not depend on the whole history $\{\hv_{\tau}\}_{\tau < t}$. In this sense, the Markovian jump process is an history-independent Poisson process, in contrast to the non-Markovian jump process (or the history-dependent Poisson process) defined as Eq.~\eqref{eq:def_history_dependent_Poisson}. 
				}
			\end{figure}
			
			Markovian jump processes constitute a large subclass of Markovian SDEs, such that 
			\begin{equation}
				\frac{\dd \hv_t}{\dd t} = \hxi^{\mrP}_{\lambda(y|\hv_t)},
				\label{eq:review:Markovian_Jump}
			\end{equation}
			where the drift term and the white Gaussian noise term are absent, and only the jump term is present. Markovian jump processes depend only on the current state $\hv_t$ and can be called {\it history-independent Poisson processes}, in contrast to the non-Markovian jump processes (or history-dependent Poisson processes) defined in Eq.~\eqref{eq:def_history_dependent_Poisson} in the main section. 

			Markovian jump processes are popular models. For instance, the detailed description of physical Brownian motion is often modelled as a Markovian jump process for the velocity of the Brownian particle, where the velocity discontinuously changes due to molecular collisions. 
		
		\subsection{The system-size expansion}
			Solving the eigenvalue problem is in general difficult, in particular when the linear operator $\mcL$ leads to an integro-differential equation of the form (\ref{eq:review:ME_standardForm}). One of the systematic methods to obtain asymptotic solutions was invented by van Kampen, which is called the {\it system-size expansion}. Mathematically, the system-size expansion can be regarded as a weak-noise asymptotic limit for Markovian jump processes. This assumption is very natural particularly in the context of Brownian motions. This method provides a solid mathematical derivation of the Langevin equations from microscopic physical dynamics. In this subsection, we briefly review this methodology based on Refs.~\cite{KzBook,KzPRL,KzJStat}. 
			
			\subsubsection{Sketch of the system-size expansion}	
				\begin{figure*}
					\centering
					\includegraphics[width=175mm]{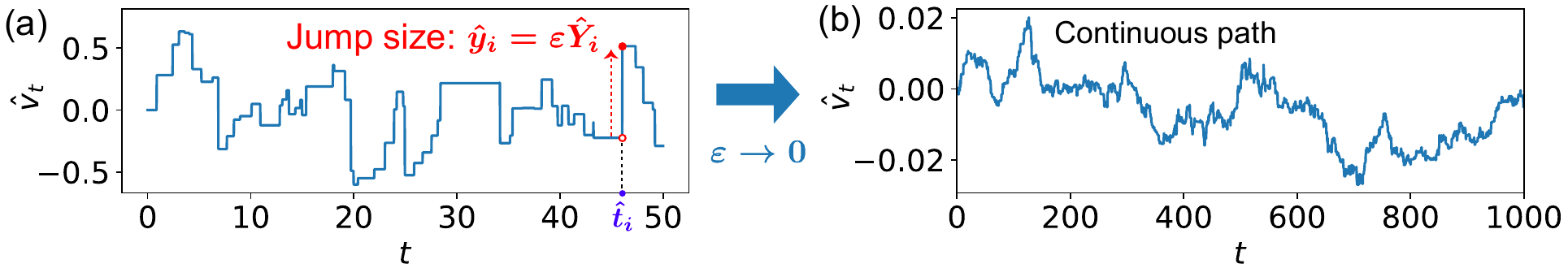}
					\caption{
						Schematic of the system-size expansion. (a)~A typical trajectory of a Markovian jump process~\eqref{eq:review:Markovian_Jump_for_SSE} where the jump size is scaled with $\ve$, such that $\hy_i=\ve \hY_i$ with $\ve$-independent jump size $\hY_i$. (b)~For small $\ve \ll 1$, the path becomes approximately continuous due to the small jump sizes and obeys the Langevin equation~\eqref{eq:review:Langevin} under an appropriate stability condition around $\hv_t\simeq 0$. This picture essentially applies even to the non-Markovian jump process~\eqref{eq:def_history_dependent_Poisson} as shown in Sec.~\ref{sec:system_size_expansion}. 
					}
					\label{fig:SSE}
				\end{figure*}
				Let us consider a Markovian jump process described by 
				\begin{equation}
					\frac{\dd\hv_t}{\dd t} = \hxi^{\mrP}_{\lambda(y|\hv_t)}, \>\>\> \hxi^{\mrP}_{\lambda(y|\hv_t)} = \sum_{k=1}^{\hN(t)}\hy_k \delta(t-\htt_k),
					\label{eq:review:Markovian_Jump_for_SSE}
				\end{equation}
				where $\lambda(y|\hv_t)$ is the conditional intensity density of the jump size $y$, $\hN(t)$ is the total number of jumps during $[0,t)$, and $\htt_k$ is the $k$-th jump time, $\hy_k$ is the $k$-th jump size. 
				
				Let us assume that the jump size is proportional to a small positive parameter $\ve>0$ (see Fig.~\ref{fig:SSE}(a)), so that we can write 
				\begin{equation}
					\hy_k = \ve \hY_k~.   
					\label{ethbgq7}
				\end{equation}
				This implies that the noise term can be rewritten as $\hxi^{\mrP}_{\lambda(y|\hv_t)} = \ve \hxi^{\mrP}_{W(Y|\hv_t)}$ with the conditional intensity density $W(Y|\hv_t)$ for the rescaled jump size $Y$. We thus obtain the SDE with a small jump-noise term: 
				\begin{equation}
					\frac{\dd \hv_t}{\dd t} = \eps \hxi^{\mrP}_{W(Y|\hv_t)}~.
					\label{eq:review_SSE_Markov_scaling}
				\end{equation}
				This is the scaling assumption for the system-size expansion. In other words, the small parameter $\eps$ can be interpreted as the small constant quantifying the weak-coupling with the stochastic environment. 

				In the small-noise asymptotic limit $\ve \to 0$ and for a broad variety of setups, assuming a stability condition around $\hv_t\simeq 0$ (see Appendix~\ref{sec:app:review_SSE} for details),
				the Markovian jump process reduces to the Langevin equation (see Fig.~\ref{fig:SSE}(b) for a schematic)
				\begin{equation}
					\frac{\dd \hv_t}{\dd t} = -\gamma \hv_t + \sqrt{2\gamma T}\hxi^{\mrG},
					\label{eq:review:Langevin}
				\end{equation}
				where $\gamma$ takes the meaning of a frictional constant and $T$ is the temperature. See Appendix~\ref{sec:app:review_SSE} for the detailed derivation. Thus, the system-size expansion is a celebrated mathematical foundation for the derivations of the Langevin equations from microscopic dynamics. 

			\subsubsection{Physical validity of the scaling assumption.}
				\begin{figure*}
					\centering
					\includegraphics[width=165mm]{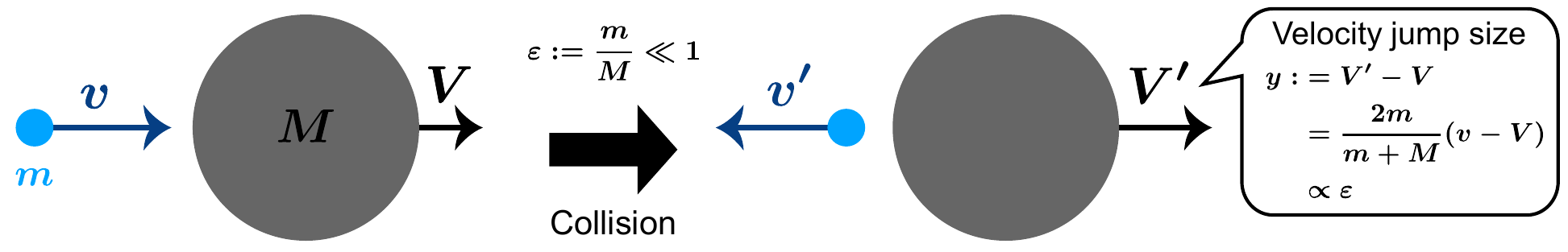}
					\caption{
						The scaling assumption of the system-size expansion is natural on physical grounds for the description of a massive Brownian-particle motion. Let us denote the mass and initial velocity of the Brownian particle (respectively of a surrounding small particle) by $M$ (respectively $m$) and $\hV$ (respectively $\hv$). After an elastic collision, the velocity of the Brownian particle changes according to formula~\eqref{eq:review:velocity_after_collision} deriving from the conservation of momentum. The velocity jump size $y$ is proportional to the mass ratio $\ve := m/M$, which is the small parameter of the problem in the limit of massive Brownian particle limit $m\ll M$. 
					}
					\label{fig:review:MassiveBrownian}
				\end{figure*}
				The physical validity of the scaling \eqref{ethbgq7} and  \eqref{eq:review_SSE_Markov_scaling} can be intuitively understood by considering a one-dimensional collision problem (Fig.~\ref{fig:review:MassiveBrownian}). Let us prepare a small particle of mass $m$, velocity $v$ and a large particle of mass $M$ and velocity $V$. In a one-dimensional elastic collision, the post-collisional velocity $V'$ of the large particle is given by 
				\begin{equation}
					V' - V = \frac{2m}{m+M}(v-V). 
					\label{eq:review:velocity_after_collision}
				\end{equation}
				Since the typical thermal velocities are given by $v\simeq \sqrt{T/m}$ and $V\simeq \sqrt{T/M}$ where $T$ is the gas temperature, 
				we have $|V/v| \propto \eps^{1/2} \ll 1$ with $\eps :=m/M$. For $\eps \ll 1$, we obtain the velocity jump $y$ of the large particle as 
				\begin{equation}
					y:= V' - V \simeq 2\eps v. 
				\end{equation}
				Thus, the velocity-jump size is proportional to $\ve$, and satisfies the system-size expansion scaling~\eqref{eq:review_SSE_Markov_scaling} exactly. This example highlights that the scaling assumption of the system-size expansion is physically reasonable\footnote{Note that the assumption that the dynamics is Markovian is also valid for Brownian dynamics in the dilute-gas limit.} when the Brownian particle in a gas is much heavier than the surrounding gas particles. 

			\subsubsection{Scaling assumption in the master equations.}
				The scaling assumption $y=\ve Y$ at a trajectory level is equivalent to the scaling assumption for the ME:
				\begin{equation}
					\lambda(y|v) = \frac{1}{\ve}W\left(\frac{y}{\ve} \Big| v\right),
				\end{equation}
				which is derived from the conservation of probability (i.e., the Jacobian relation), such that $\lambda(y|v)dy = W(Y|v)dY$.
				
		\subsection{Goal of this report}
				On the basis of the above theory regarding the standard forms of SDEs and MEs, our goals in this report are the following. 
				\begin{enumerate}
					\item We derive the ME for the general non-Markovian jump process analogous to the standard form of the Markovian ME~\eqref{eq:review:ME_standardForm}.
					\item We asymptotically solve the ME for non-Markovian jump processes by generalising the system-size expansion; we finally obtain the GLE via a physically-reasonable coarse-graining approach. 
				\end{enumerate}

\section{Non-Markovian model and formulation}\label{sec:Model}
	In this section, we first present the stochastic model studied in this report. We then introduce the Laplace-convolution Markovian embedding 
	that converts the original low-dimensional non-Markovian dynamics onto a Markovian field dynamics. Finally, the corresponding field ME is formulated. 

	\subsection{Non-Markovian jump process (history-dependent Poisson process)}
		\begin{figure}
			\centering
			\includegraphics[width=85mm]{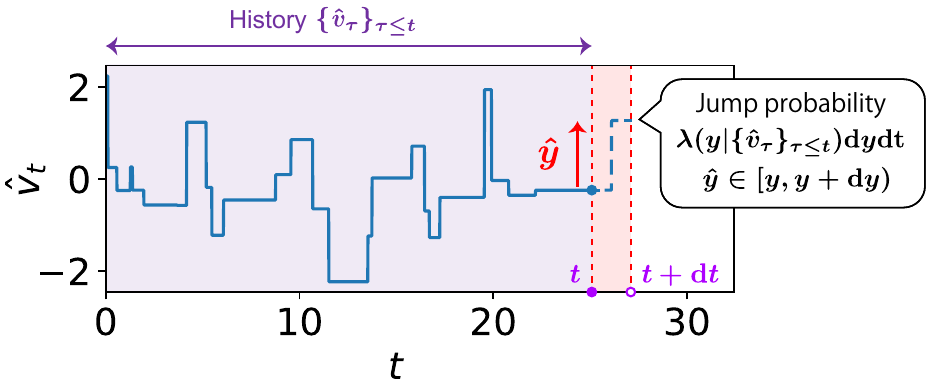}
			\caption{
				Schematic of the non-Markovian jump process (or history-dependent Poisson process). The probability that a velocity jump occurs during $[t,t+\dd t)$ is given by $\lambda(y|\{\hv_{\tau}\}_{\tau\leq t})dt dy$ with jump size $\hy \in [y, y+\dd y)$. Here the intensity density $\lambda(y|\{\hv_{\tau}\}_{\tau\leq t})$ explicitly depends on the whole history $\{\hv_{\tau}\}_{\tau\leq t}$ and the system is thus truly non-Markovian. 
			}
			\label{fig:main:non-Markov_path}
		\end{figure}
		Let us consider a non-Markovian stochastic model that can encompass a large class of non-Markovian stochastic processes:
		\begin{box_normal}
			We study the history-dependent compound Poisson process (see 
			Fig.~\ref{fig:main:non-Markov_path} for a schematic)
			\begin{equation}\label{eq:def_history_dependent_Poisson}
				\frac{\dd \hv_t}{\dt} = \hxiCP_{\lambda(y|\{\hv_\tau\}_{\tau\leq t})},
			\end{equation}
			where the intensity $\lambda(y|\{\hv_\tau\}_{\tau\leq t})$ with jump size $y$ is conditional on the full history of the system $\{\hv_\tau\}_{\tau\leq t}$. 
		\end{box_normal}
		The non-Markovian nature of this process makes the intensity a functional of the whole history. More technically, 
		Eq.~\eqref{eq:def_history_dependent_Poisson} implies that
		\begin{equation}
			\dd \hv_t := 	\begin{cases}
							\hat{y} +O(\dt)& (\mbox{probability} = \dt \dy\lambda(y|\{\hv_\tau\}_{\tau\leq t}) \mbox{ for any } \hat{y} \in [y,y+\dy)) \\
							0 & (\mbox{probability} = 1 - \dt\int_{-\infty}^\infty \dy\lambda(y|\{\hv_\tau\}_{\tau\leq t}))
						\end{cases}
		\end{equation}
		for any given history $\{\hv_\tau\}_{\tau\leq t}$ with infinitesimal time evolution $\dd \hv_t := \hv_{t+\dt} - \hv_t$. Our aim is to provide the full analytical toolset for this history-dependent Poisson process by developing the corresponding field ME and by analysing its asymptotic solutions. 

	\subsection{Markovian embedding}
		In this subsection, we apply the Markovian-embedding scheme to the history-dependent Poisson process~\eqref{eq:def_history_dependent_Poisson}. We finally obtain the SPDE govering the Markovian field dynamics and derive the corresponding field ME. 

		\subsubsection{Basic idea}
			The idea of Markovian embedding is very simple: a low-dimensional non-Markovian dynamics can be converted onto a higher-dimensional Markovian dynamics by adding a sufficient number of auxiliary variables. This approach dates back to Mori~\cite{MoriEmbedding} around the mid 1960s\footnote{He proposed a systematic expansion of the relaxation memory kernel by the continued-fraction expansion. Truncating the expansion leads to an approximation based on the sum of several exponential memories.}. Also, the theory of the Kac-Zwanzig model~\cite{ZwanzigTB,Kac1965,Kac1987,Zwanzig1980,KupfermanJSP2004} can be regarded a theory of Markovian embedding between the generalised Langevin equation and the Hamiltonian-particles model with harmonic interaction. For example, the generalised Langevin equation with the sum of $K$-exponential memories can be thought of as a $K$-dimensional Markovian dynamics~\cite{ZwanzigTB,KupfermanJSP2004,KzDidier2019PRR}. This idea can be even applied to the Hawkes processes~\cite{BouchaudTradebook2018,Dassios2013,Hainaut2022} for memory kernels expressed as a sum of exponential functions. Remarkably, this idea of Markovian embedding has been also applied to non-Markovian stochastic processes in quantum systems~\cite{Imamoglu1994,Garraway1996,Garraway1997,Tamascelli2018,Tamascelli2019,Teretenkov2019,Pleasance2020} in the context of the pseudomode approach around the mid 1990s. 

			The dimension needed for the Markovian embedding depends on the model but can be infinite in general. In this case, the dynamics can be regarded as a Markovian field dynamics. For instance, the GLE and the Hawkes processes have been converted onto Markovian field dynamics~\cite{KzDidier2019PRL,KzDidier2019PRR,KzDidier2021PRL,KzDidier2023PRR}, which can be analysed by the field ME (which is a functional-differential equation for the probability density functional). 

			Markovian embedding is nontrivial and technically tricky for continuous-time stochastic processes, while Markovian embedding is rather straightforward for discrete-time stochastic processes (see Appendix~\ref{sec:app:DiscreteTime_Embedding} for brief clarification). This report aims at formulating a general embedding theory of the non-Markovian jump process~\eqref{eq:def_history_dependent_Poisson}, even though it is based on continuous time.

		\subsubsection{Variable set}
			\begin{figure*}
				\centering
				\includegraphics[width=130mm]{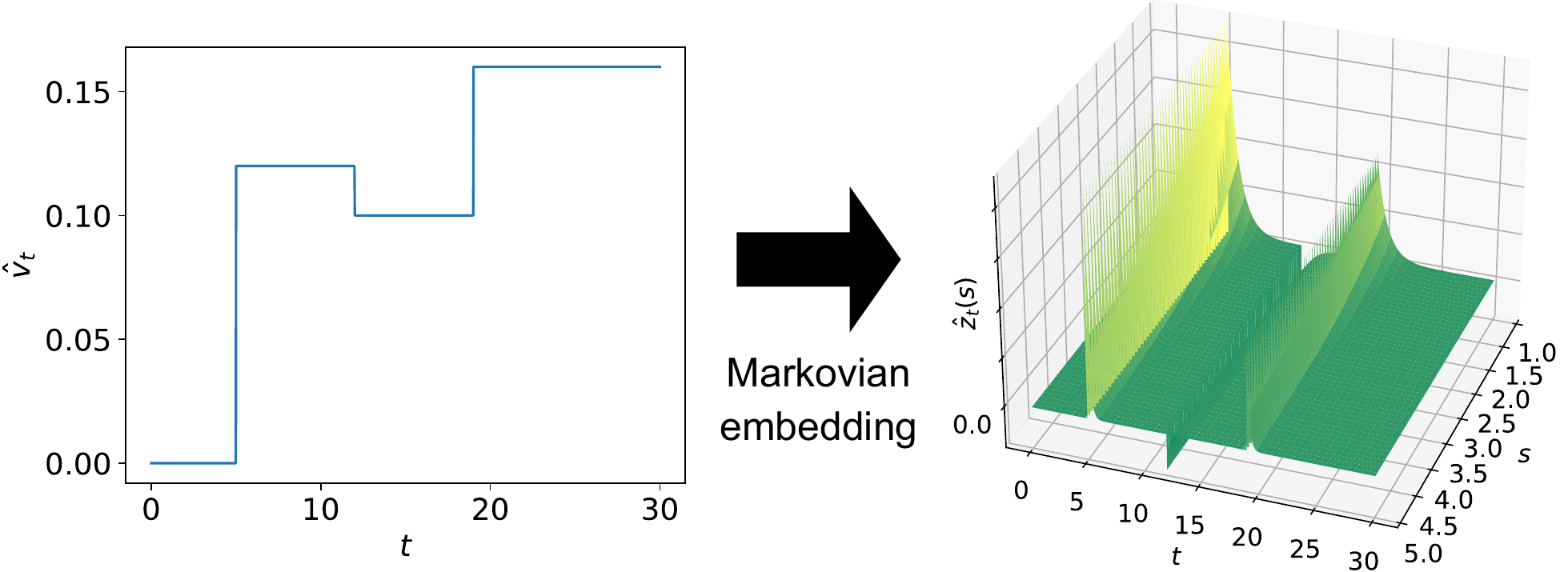}
				\caption{
					Schematic of the Markovian embedding of the original one-dimensional non-Markovian jump process~\eqref{eq:def_history_dependent_Poisson} onto the Markovian field dynamics $\{\hz_t(s)\}_s$. The auxiliary field variables $\{\hz_t(s)\}_s$ are defined by Eq.~\eqref{def:Laplace_MarkovEmbedding_HDP} on the wave-number axis $s \in (0,\infty)$ (i.e., one-dimensional field) and obey the first-order Markovian SPDE~\eqref{eq:set_complete_dynamics_hdCP}.
				}
				\label{fig:MarkovianEmbedding_AuxiliaryField}
			\end{figure*}
			Before proceeding with the derivation of the Markovian field dynamics, let us introduce a complete set of system variables useful for Markovian embedding. In the previous section, we used $\{\hv_{\tau}\}_{\tau< t}$ as a naive complete set of system variables. This set is equivalent in information content to another set $(\hv_t;\{\hv_\tau \}_{\tau\leq t})$ with acceleration $\hta_t:=\dd \hv_t/\dt$. Note that the acceleration can include the impulses described by the Dirac $\delta$ functions associated with the jumps. Let us now introduce the {\it Laplace-convolution Markovian-embedding representation} of the velocity trajectory as
			\begin{box_normal}
				\begin{equation}
					\label{def:Laplace_MarkovEmbedding_HDP}
					\hz_t(s) := \int_0^\infty e^{-s\tau}\hta_{t-\tau}\dd \tau, \>\>\> 
					\hta_t := \frac{\dd \hv_t}{\dt},
				\end{equation}
				which is defined for $s>0$ (see Fig.~\ref{fig:MarkovianEmbedding_AuxiliaryField} for a schematic). We then adopt the variable set
				\begin{equation}
					(\hv_t, \{\hz_t(s)\}_{s>0})
				\end{equation}
				as a useful complete variable set.
			\end{box_normal}
			The introduction of the auxiliary field variable $\{\hz_t(s)\}_{s>0}$ is the technical but crucial trick to convert the general nonlinear non-Markovian model onto a Markovian field model. In the following, the wave number $s$ is always considered strictly positive ($s>0$), and the set $\{f(s)\}_{s>0}$ for any function $f(s)$ is sometimes abbreviated by $\{f(s)\}_{s}$ if its meaning is clear from the context. 
		
		\subsubsection{Phase space}
			Let us introduce the state variables $\hGamma_t$ as points in the phase space $\bm{S}$, such that 
			\begin{equation}
				\hGamma_t := (\hv_t, \{\hz_t(s)\}_{s}) \in \bm{S}, \>\>\> \bm{S}:= \{(v, \{z(s)\}_{s}) : v \in \bm{R}, z(s) \in \bm{F} \}
			\end{equation}
			where $\bm{R}$ is the space of real numbers and $\bm{F}$ is the function space. 
			In the following, we simplify the notation of functionals such as the intensity as a functional in terms of the history in the following way:
			\begin{equation}
				\lambda[y|\hGamma_t] := \lambda [y| \hv_t; \hz_t] :=  \lambda(y|\hv_t,\{\hta_\tau\}_{\tau\leq t}) := \lambda(y|\{\hv_\tau\}_{\tau\leq t})
			\end{equation}
			In our notation, the functional argument (e.g., $\hGamma_t$ and $\{\hz_t(s)\}_{s}$) follows other variables (e.g., $y$ and $\hv_t$) after the separation by
			the semi-colon. 

			Note that the ordinary Markov compound Poisson process corresponds to the case where the intensity does not depend on the historical velocities, such that
			\begin{equation}
				\lambda[y|\hv_t;\hta_t] = \lambda(y|\hv_t)
			\end{equation}
			with a nonnegative function $\lambda(y|\hv_t)$. 
		
	\subsection{Markovian field dynamics}
		The history-dependent compound Poisson process~\eqref{eq:def_history_dependent_Poisson} characterising the original variables $\{\hv_\tau\}_{\tau\leq t}$ is equivalent to the set of the following SDE and SPDEs characterising the new variables $\hGamma_t:=(\hv_t,\{\hz_t(s)\}_s)$: 
		\begin{box_normal}
			\begin{subequations}
				\label{eq:set_complete_dynamics}
			\begin{align}\label{eq:set_complete_dynamics_hdCP}
				\frac{\dd \hv_t}{\dt} = \hxiCP_{\lambda[y|\hGamma_t]}, \>\>\>
				\frac{\partial \hz_t(s)}{\partial t} = -s\hz_t(s) + \hxiCP_{\lambda[y|\hGamma_t]},
			\end{align}
			where the jump term $\hxiCP_{\lambda[y|\hGamma_t]}$ simultaneously acts on both $\hv_t$ and $\hz_t(s)$ for all $s>0$ (see Fig.~\ref{fig:MarkovianEmbedding_AuxiliaryField} for a typical configuration of the auxiliary field). The initial condition is given by 
			\begin{equation}
				z_0(s) = \int_0^\infty e^{-s\tau}\hta_{-\tau}\dd \tau, \>\>\> 
				\hta_\tau := \frac{\dd \hv_\tau}{\dd \tau}.
			\end{equation}
			This set of SDEs characterizes the complete dynamics of the phase point $\hGamma_t=(\hv_t, \{\hz_t(s)\}_{s})$ in a closed form. 				
			\end{subequations}
		\end{box_normal}

		The essence of the trick is to use the Laplace-convolution transform, which encodes the whole history of $\hv_t$ (or equivalently its acceleration $\hta_t$) into a function of $t$ (now, thus Markovian) and of an additional variable $s$. The function $\hz_t(s)$ dependent on $s$ serves as the key device to render the system Markovian, utilizing an infinite series of equations for all $\hz_t(s)$. It is remarkable that this system is Markovian in the extended phase space $\hGamma_t\in \bm{S}$, while the original one-dimensional process is non-Markovian. This means that we have successfully transformed the original non-Markovian dynamics into a Markovian dynamics by adding a sufficient number of variables. Since the resulting dynamics is Markovian, we can derive the corresponding ME for the PDF for the phase point $\hGamma_t$ in the extended phase space. 

		\subsubsection*{Derivation}
			By directly solving Eq.~\eqref{eq:set_complete_dynamics_hdCP}, we obtain 
			\begin{align}
				\hz_t(s) &= \hz_0(s)e^{-st} + \int_{0}^t e^{-s(t-t')}\hxiCP_{t',\lambda[y|\hGamma_t]}\dt' \notag \\ 
				&= \int_0^\infty e^{-s(\tau+t)}\hta_{-\tau}\dd \tau + \int_{0}^t e^{-s(t-t')}\hta_{t'}\dt' \notag \\ 
				&= \int_t^\infty e^{-s\tau'}\hta_{t-\tau'}\dd \tau' + \int_{0}^t e^{-st''}\hta_{t-t''}\dt''\notag \\
				&= \int_0^\infty e^{-s\tau}\hta_{t-\tau}\dd \tau
			\end{align}
			with the dummy-variable transformation $\tau':=t+\tau$ and $t'':=t-t'$ from the second to the third line. Thus, the set of the SDEs~\eqref{eq:set_complete_dynamics_hdCP} is consistent with the Markovian-embedding representation~\eqref{def:Laplace_MarkovEmbedding_HDP}.

	\subsection{Field master equation for the history-dependent Poisson process}
		The functional ME of the field corresponding to the SPDEs~\eqref{eq:set_complete_dynamics_hdCP} are given by 
		\begin{box_normal}			
			\begin{equation}\label{eq:field_master_hdCP}
				\frac{\partial P_t[\Gamma]}{\partial t} = \mathcal{L}P_t[\Gamma] := \int_0^\infty \ds \frac{\delta}{\delta z(s)}\left(sz(s)P_t[\Gamma]\right) + 
				\int_{-\infty}^{\infty}\dy\left\{\lambda[y|\Gamma-\Delta \Gamma_y]P_t[\Gamma-\Delta \Gamma_y]-\lambda[y|\Gamma]P_t[\Gamma]\right\}.
			\end{equation}
			with jump size vector
			\begin{equation}
				\Delta \Gamma_y := (y, \{y\bm{1}(s)\}_{s})
			\end{equation}
			with indicator function $\bm{1}(s)=1$ for any $s$. 
		\end{box_normal}

		\subsubsection*{Derivation}
			We derive the field ME as follows.
			For any functional $f[\Gamma_t]:= f[\hv_t;\hz_t] = f(\hv_t, \{\hz_t(s)\}_{s})$, from equation (\ref{eq:set_complete_dynamics_hdCP}), its path-level differential $\dd f[\hGamma_t]:= f[\hGamma_{t+dt}]-f[\hGamma_t]$ is given by
			\begin{equation}
				\df[\hGamma_t] = \begin{cases}
									f[\hGamma_t+\Delta \Gamma_{\hy}] - f[\hGamma_t] + O(\dt) & (\mbox{prob.} = \dt \dy\lambda[y|\hGamma_t]\mbox{ for any }\hy \in [y,y+\dy)) \\
									\displaystyle -\dt \int_0^\infty s\hz_t(s)\frac{\delta f[\hGamma_t]}{\delta \hz_t(s)}\ds + O(\dt^2)& (\mbox{prob.} = 1-\dt\int_{-\infty}^\infty \dy\lambda[y|\hGamma_t])
								\end{cases}
			\end{equation}
			at leading order\footnote{Notably, while the jump probability during $[t,t+dt)$ is of order $dt$, the leading-order contribution $f[\hGamma_t+\Delta \Gamma_{\hy}] - f[\hGamma_t]$ is of order $1$. In addition, while the no-jump probability during $[t,t+dt)$ is of order $1$, the leading-order contribution $-\dt \int_0^\infty s\hz_t(s)\{\delta f[\hGamma_t]/\delta \hz_t(s)\}\ds$ is of order $dt$. Therefore, the contributions of their averages are balanced at the order $dt$.}. By taking the ensemble average of both sides, up to the order of $\dt$, we obtain
			\begin{align}
				\left< \df[\hGamma_t]\right> = \dt \int \dd \Gamma P_t[\Gamma]\left[\int_{-\infty}^\infty \dd y\lambda[y|\Gamma]\left(f[\Gamma+\Delta \Gamma_y]-f[\Gamma]\right) 
				-  \int_0^\infty \ds\left\{sz(s)\frac{\delta }{\delta z(s)}f[\Gamma]\right\}\right] + O(\dt^2)
				\label{eq:ME_der_transform}
			\end{align}
			with integral volume element $\dd\Gamma:= \dd v \mcD z$. By applying a variable transformation $\Gamma+\Delta \Gamma_y = \Gamma'$,  the first term in the right-hand side  is given by 
\begin{equation}
\int \dd \Gamma P_t[\Gamma]\lambda[y|\Gamma]f[\Gamma+\Delta \Gamma_y] 
				=\int \dd \Gamma' P_t[\Gamma'-\Delta \Gamma_y]\lambda[y|\Gamma'-\Delta \Gamma_y]f[\Gamma'] 
				=\int \dd \Gamma P_t[\Gamma-\Delta \Gamma_y]\lambda[y|\Gamma-\Delta \Gamma_y]f[\Gamma],
\end{equation}	
			where the dummy variable $\Gamma'$ is finally replaced with $\Gamma$. By applying the functional partial integration~\eqref{eq:app:partialIntegral_functional}, the third term in the right-hand side of Eq.~\eqref{eq:ME_der_transform} is given by 
			\begin{equation}
				\int  sz(s)P_t[\Gamma]\frac{\delta f[\Gamma]}{\delta z(s)}\dd \Gamma
				= - \int  f[\Gamma] \frac{\delta}{\delta z(s)}\left\{sz(s)P_t[\Gamma]\right\}\dd \Gamma.
			\end{equation} 
			By considering that the left-hand side of Eq.~\eqref{eq:ME_der_transform} is given by
			\begin{equation}
				\left< \df[\hGamma_t]\right> = 
				\left< f[\hGamma_{t+dt}]\right> - \left< f[\hGamma_{t}]\right> = 
				\int \dd\Gamma\left(P_{t+\dt}[\Gamma]-P_t[\Gamma]\right)f[\Gamma] = \dt\int \dd\Gamma f[\Gamma]\frac{\partial P_{t}[\Gamma]}{\partial t} + O(\dt^2),
			\end{equation}
			we finally obtain the following integral identity regarding any functional $f[\Gamma]$
			\begin{align}
				\int \dd\Gamma f[\Gamma]\frac{\partial P_t[\Gamma]}{\partial t} = \int d\Gamma f[\Gamma]\left[\int_{-\infty}^\infty \dy\left\{\lambda[y|\Gamma-\Delta \Gamma_y]P_t[\Gamma-\Delta \Gamma_y]-\lambda[y|\Gamma]P_t[\Gamma]\right\} 
				+ \int_0^\infty \ds\frac{\delta}{\delta z(s)}\left(sz(s)P_t[\Gamma]\right)\right]
			\end{align}
			in the limit $\dt\to 0$. Since this relation holds for an arbitrary $f[\Gamma]$, we obtain the field ME~\eqref{eq:field_master_hdCP}. 

	\subsection{Functional Kramers-Moyal expansion}
		By applying the identity (see Eq.~\eqref{eq:app:fullTaylorfunctional_2var} for the functional Taylor expansion)
		\begin{equation}
			\int_{-\infty}^{\infty}\dy\lambda[y|\Gamma-\Delta \Gamma_y]P_t[\Gamma-\Delta \Gamma_y] = \sum_{n=0}^\infty \frac{1}{n!}\int_{-\infty}^\infty \dy y^n \left(-\frac{\partial }{\partial v}-\int_0^{\infty} \ds\frac{\delta}{\delta z(s)}\right)^n\lambda[y|\Gamma]P_t[\Gamma],
		\end{equation}
		\begin{subequations}
			to the field ME~\eqref{eq:field_master_hdCP}, we obtain the functional Kramers-Moyal (KM) expansion: 
			\begin{box_normal}
			\begin{equation}
				\frac{\partial P_t[\Gamma]}{\partial t} = \int_0^\infty \ds\frac{\delta }{\delta z(s)}\left(sz(s)P_t[\Gamma]\right) + \sum_{n=1}^\infty \frac{(-1)^n}{n!}\left(\frac{\partial }{\partial v}+\int_0^{\infty} \ds\frac{\delta}{\delta z(s)}\right)^n\alpha_n[\Gamma]P_t[\Gamma]
				\label{eq:KM_expansion_formula}
			\end{equation}
			with the KM coefficient
			\begin{equation}
				\alpha_n[\Gamma]:=\int_{-\infty}^\infty y^n\lambda[y|\Gamma]\dy.
			\end{equation}
			\end{box_normal}
		\end{subequations}

	\subsection{Remark: systematic calculations based on linear algebra}
		Since the field ME~\eqref{eq:field_master_hdCP} is linear, it can be analysed with the tools of linear algebra. 
Starting from the standard form (\ref{eq:review:ME}) $\frac{\pd P_t[\Gamma]}{\pd t} = \mcL P_t[\Gamma]$, let us consider
the eigenvalue problem 
		\begin{equation}
			\mathcal{L}\phi_i[\Gamma] = -\mu_i \phi_i [\Gamma]
		\end{equation} 
		where $\mu_i$ is the $i$-th eigenvalue and $\phi_i[\Gamma]$ is the corresponding eigenfunction\footnote{While the eigenvalue spectral may be continuous technically, we formally write the eigenvalues with discrete notation.}. 
		The time-dependent solution is then given by superposition of the eigenfunctions
		\begin{equation}\label{eq:time-dependent_sol_eigen}
			P_t[\Gamma] = \sum_{i} c_i e^{-\mu_i t}\phi_i[\Gamma],
		\end{equation}
		where the coefficients $\{c_i\}_i$ are determined by the initial condition. 
		The steady-state PDF corresponds to the zeroth eigenfunction $\phi_i [\Gamma]$ with $\mu_0 = 0$:
		\begin{equation}\label{eq:steady-state_sol_eigen}
			P_{\mrss}[\Gamma] \propto \phi_0[\Gamma].
		\end{equation} 
		Various physical quantities can be systematically calculated by the time-dependent~\eqref{eq:time-dependent_sol_eigen} or steady-state solutions~\eqref{eq:steady-state_sol_eigen}. For example, the correlation function is formally given by the path integral 
		\begin{equation}
			\la \hat{A}_t \hat{B}_{t'}\ra = \int  A_t[\Gamma] B_{t'}[\Gamma] P_{\mrss}[\Gamma]\dd \Gamma
		\end{equation}
		where $A_t$ and $B_{t'}$ are expressed as functionals of $\Gamma$.

\section{Application 1: System-size expansion and generalized Langevin equation}\label{sec:system_size_expansion}
In this section, we illustrate the utilization of our field ME framework in relation to an asymptotic theory, drawing upon the system-size expansion. 
By enforcing a stability condition upon the system-size expansion, we ultimately infer the Generalized Langevin Equation (GLE) as a plausible coarse-graining process rooted in physical reasoning.

	\subsection{Assumptions \label{thbtbvqag}}
	\begin{box_normal}{}
	\begin{enumerate}
		\item 	{\bf Small noise assumption:} 
				Let us consider the non-Markovian process with small jumps:  
				\begin{equation}
					\frac{\dd \hv_t}{\dd t} = \ve \hxi^{\mrP}_{W[Y|\{\hv_{\tau}\}_{\tau\leq t}]},
				\end{equation}
				where $W$ is the $\ve$-independent conditional intensity of the jump size $Y$ (see Fig.~\ref{fig:SSE} for the schematic of the small-jump assumption). 
				This assumption is equivalent to the following scaling relation of the conditional intensity density 
				\begin{equation}
					\lambda[y|\Gamma] = \frac{1}{\ve}W\left[\frac{y}{\ve}\bigg|\Gamma\right].
				\end{equation}
				This assumption leads to the scaling relation for the KM coefficients
				\begin{equation}
					\alpha_n[\Gamma] = \ve^n \mcA_n[\Gamma], \>\>\> \mcA_n[\Gamma]:=\int_{-\infty}^\infty Y^n W[Y|\Gamma]\dd Y
				\end{equation}
				with the $\ve$-independent KM coefficient $\mcA_n[\Gamma]$. 
	
		\item 	{\bf Linear stability:}
				Let us additionally assume that the first-order KM coefficient $\mcA_1[\Gamma]$ has a single stable point:
				\begin{equation}
					\mcA_1[\Gamma=\bm{0}] = 0,\>\>\> 
					\bm{0} := (0,\{0\}_{s})
				\end{equation}
				We also assume that $\mcA_1[\Gamma]$ is linearly stable around $\Gamma=\bm{0}$, such that 
				\begin{subequations}
					\begin{align}\label{def:gamma}
						\gamma = - \frac{\partial }{\partial v}\mcA_1[\Gamma]\bigg|_{\Gamma = \bm{0}} > 0,  \>\>\>
						\Upsilon(u) = -\frac{\delta }{\delta z(s)}\mcA_1[\Gamma]\bigg|_{\Gamma = \bm{0}} > 0 \>\> \mbox{for any }u
					\end{align}
				\end{subequations}
				where the rescaled wave number $u$ is defined by $u:=\frac{s}{\ve}$.
				
		\item {\bf Existence of the noise term:}
				The noise term is assumed to be present even for $\ve \to 0$ and thus the variance term is nonzero: 
				\begin{equation}
					\sigma^2 = \mcA_2[\Gamma=\bm{0}] > 0. 
				\end{equation}
	\end{enumerate}
	\end{box_normal}

	As a technical assumption, we assume that all the considered integrals converge. This assumption implies that physically singular processes, such as with long time tail with decaying speed slower than $t^{-1}$, are out of the scope of this paper. Note that the above stability assumptions parallel the conventional stability assumption for the Markovian jump process (see Appendix~\ref{sec:app:review_SSE} for their comparison). Also, we note that all even-order KM coefficients are positive under assumption 3, such that $\mcA_k[\bm{0}] > 0$ for all even $k$ due to Pawula's theorem~\cite{RiskenB}.
	
	\subsection{Asymptotic derivation of the functional Fokker-Planck equation}
		\begin{box_normal}
			Given the following rescaled variables
			\begin{equation}
				\mft := \ve t, \>\>\> u := \frac{s}{\ve}, \>\>\> V := \frac{v}{\sqrt{\ve}}, \>\>\> Z(u) := \frac{z(s)}{\sqrt{\ve}}, \>\>\> G:= (V,\{Z(u)\}_{u}),
			\end{equation}
			we obtain the following functional Fokker-Planck equation for $\ve \to 0$
			\begin{equation}
				\frac{\partial P_{\mft}[G] }{\partial \mft}
				= \left[\gamma\frac{\partial }{\partial V} V 
				+ \int_0^\infty \dd u_1\frac{\delta }{\delta Z(u_1)} \left\{
					uZ(u_1)  +\int_0^{\infty} \dd u_2 \Upsilon(u_2)Z(u_2)
				\right\} 
				+ \frac{\sigma^2}{2}\left(\frac{\partial }{\partial V}+\int_0^{\infty} \dd u\frac{\delta}{\delta Z(u)}\right)^2 \right]P_{\mft}[G]. 
				\label{eq:FP_functional_SSE}
			\end{equation}
		\end{box_normal}

		\subsubsection{Derivation}
			With the above assumptions, let us formulate the system size expansion for this model. We have 
			\begin{align}
				\int_0^\infty \ds\frac{\delta }{\delta z(s)}\left(sz(s)P_t[\Gamma]\right)
				= \ve\int_0^\infty \dd u\frac{\delta }{\delta Z(u)}\left(u Z(u)P_t[\Gamma]\right),
			\end{align}
			and 
			\begin{align}
				\left(\frac{\partial }{\partial v}+\int_0^{\infty} \ds\frac{\delta}{\delta z(s)}\right)^n\alpha_n[\Gamma]P_t[\Gamma]
				&= \left(\ve^{-1/2}\frac{\partial }{\partial V}+\ve^{-1/2}\int_0^{\infty} \dd u\frac{\delta}{\delta Z(u)}\right)^n \ve^n \mcA_n[\Gamma]P_t[\Gamma] \notag \\
				&= \ve^{n/2}\left(\frac{\partial }{\partial V}+\int_0^{\infty} \dd u\frac{\delta}{\delta Z(u)}\right)^n \mcA_n[\Gamma]P_t[\Gamma].
			\end{align}
			We also consider the functional Maclaurin series~\eqref{eq:app:Mclaurin_functional_2var} for the KM coefficient around $\Gamma=\bm{0}$ 
			\begin{align}
				\mcA_n[\Gamma] &= \mcA_n[\bm{0}] + \sum_{k=1}^\infty \frac{1}{k!}\left(v\frac{\partial }{\partial \chi}+\int_0^\infty \dd s z(s)\frac{\delta }{\delta \zeta(s)}\right)^k\mcA_n(\chi;\{\zeta(s)\}_{s})\bigg|_{(\chi,\{\zeta(s)\}_s)=\bm{0}} \notag\\
				&= \mcA_n[\bm{0}] + \sum_{k=1}^\infty \frac{\ve^{k/2}}{k!}\left(V\frac{\partial }{\partial \chi}+\int_0^\infty \dd u Z(u)\frac{\delta }{\delta \zeta(u)}\right)^k\mcA_n(\chi;\{\zeta(u)\}_{u})\bigg|_{(\chi,\{\zeta(u)\}_u)=\bm{0}}
			\end{align}
			with the dummy-variable arguments $\chi$ and $\zeta$. This relation implies that  	
			\begin{align}
				\mcA_1[\Gamma] &= \mcA_1[\bm{0}] + \sum_{k=1}^\infty \frac{\ve^{k/2}}{k!}\left(V\frac{\partial }{\partial \chi}+\int_0^\infty \dd u Z(u)\frac{\delta }{\delta \zeta(u)}\right)^k\mcA_1(\chi;\{\zeta(u)\}_{u})\bigg|_{(\chi,\{\zeta(u)\}_u)=\bm{0}} \notag \\
				&= 0 - \ve^{1/2}\left(\gamma V + \int_0^\infty \dd u \Upsilon(u) Z(u)\right) + O(\ve),
			\end{align}
			and 
			\begin{align}
				\mcA_n[\Gamma] &= \mcA_n[\bm{0}] + \sum_{k=1}^\infty \frac{\ve^{k/2}}{k!}\left(V\frac{\partial }{\partial \chi}+\int_0^\infty \dd u Z(u)\frac{\delta }{\delta \zeta(u)}\right)^k\mcA_n(\chi;\{\zeta(u)\}_{u})\bigg|_{(\chi,\{\zeta(u)\}_u)=\bm{0}} \notag \\
				&= \mcA_n[\bm{0}] + O(\ve^{1/2})~~~~{\rm for}~ n\geq 2~. 
			\end{align}
			From the KM expansion~\eqref{eq:KM_expansion_formula}, by introducing $\mft:=\ve t$, we obtain 
			\begin{align}
				\frac{\partial P_{\mft}[G]}{\partial \mft} =& \int_0^\infty \dd u\frac{\delta }{\delta Z(u)}\left(uZ(u)P_{\mft}[G]\right) + \sum_{n=1}^\infty \frac{(-1)^n\ve^{n/2-1}}{n!}\left(\frac{\partial }{\partial V}+\int_0^{\infty} \dd u\frac{\delta}{\delta Z(u)}\right)^n \mcA_n[G]P_{\mft}[G] \notag\\
				=& \int_0^\infty \dd u\frac{\delta }{\delta Z(u)}\left(uZ(u)P_{\mft}[G]\right) + 
				\left(\frac{\partial }{\partial V}+\int_0^{\infty} \dd u_1\frac{\delta}{\delta Z(u_1)}\right)\left(\gamma V + \int_0^\infty \dd u_2 \Upsilon(u_2)Z(u_2)\right)P_{\mft}[G] \notag\\
				&+ \frac{\sigma^2}{2}\left(\frac{\partial }{\partial V}+\int_0^{\infty} \dd u\frac{\delta}{\delta Z(u)}\right)^2P_{\mft}[G] + O(\ve^{1/2}). 
			\end{align}
			We thus obtain the functional Fokker-Planck equation~\eqref{eq:FP_functional_SSE} in the weak coupling limit $\ve\to 0$. 
		
		\subsubsection{Equivalent stochastic dynamics}
			Furthermore, the functional Fokker-Planck~\eqref{eq:FP_functional_SSE} is equivalent to the stochastic dynamics described by
			\begin{subequations}
				\label{eq:GaussianSPDE_SSE_trans}				
				\begin{align}
					\frac{\dd \hV_{\mft}}{\dd \mft} &= -\gamma \hV_{\mft} - \int_0^\infty \dd u\Upsilon(u)\hZ_{\mft}(u) + \sigma \hxi^{\mrG}_{\mft}, \label{eq:GaussianSPDE_SSE_trans_1}	\\
					\frac{\partial \hZ_{\mft}(u)}{\partial \mft} &= -u \hZ_{\mft}(u) -\gamma \hV_{\mft} - \int_0^\infty \dd u \Upsilon(u)\hZ_{\mft}(u) + \sigma \hxi^{\mrG}_{\mft}
					\label{eq:GaussianSPDE_SSE_trans_2}
				\end{align}
			\end{subequations}
			with standard white Gaussian noise $\hxi^{\mrG}_{\mft}$ that is common to the stochastic dynamics of $\hV_{\mft}$ and $\hZ_{\mft}(u)$.

	\subsection{Asymptotic derivation of the generalized Langevin equation}
		The stochastic dynamics~\eqref{eq:GaussianSPDE_SSE_trans} is equivalent to the generalized Langevin equation (GLE):
		\begin{box_normal}
		\begin{equation}
			\frac{\dd \hV_{\mft}}{\dd \mft} = -\gamma \hV_{\mft} -\int_{-\infty}^{\mft} \dd{\mft}'\mathcal{M}(\mft-\mft')\hX_{\mft'} + \heta_{\mft}
		\end{equation}
		with memory kernel $\mathcal{M}(\mft)$ given by expression (\ref{ghbgqb1tqa}) and colored Gaussian noise
		\begin{equation}
			\heta_{\mft} := \sigma \hxi^{\mrG}_{\mft} + \sigma \int_{-\infty}^{\mft} \dd \mft' \mathcal{M}(\mft-\mft')\hxi_{\mft'}^{\mrG}.
		\end{equation}
		\end{box_normal}
		This result implies that the GLE is a minimal model for the coarse-grained description of general non-Markovian jump processes in the weak coupling limit $\ve\to 0$ under the stability condition (\ref{thbtbvqag}).

		The memory kernel $\mathcal{M}(\mft)$ and the noise statistics can be explicitly derived as follows. Let us define the ``matrix" $K(u,u')$ 
		\begin{equation}
			K(u,u') := u\delta (u-u') + \Upsilon(u'), 
			\label{def:K_matrix}
		\end{equation}
		and the corresponding eigenvalue $\mu$ and eigenfunctions $\{e(\mu;u)\}_{\mu}$ satisfying	
		\begin{equation}
			\int_0^\infty \dd u'K(u,u')e(\mu;u') = \mu e(\mu;u).
		\end{equation}
		The matrix $K(u,u')$ has the following properties (see Appendix~\ref{sec:app:eigenvalues}): 
			(i)~All of its eigenvalues are real and positive $\mu > 0$.
			(ii)~$K(u,u')$ is a positive symmetric matrix and thus has an inverse matrix $K^{-1}(u,u')$.
		We assume that the eigenfunctions $\{e(\mu;u)\}_{\mu}$ constitute a complete set, and have inverse matrices $e^{-1}(u;\mu')$ such that \footnote{In $N$-dimensional linear algebra, the set of all eigenvectors $\{\bm{e}_i\}_i$ with $\bm{e}_i= (e_{i1},\dots, e_{iN})$ of a symmetric matrix constitute a complete set. In addition, the matrix $A:=(e_{ij})$ has the inverse matrix $A^{-1}=(e^{-1}_{ij})$, such that $\sum_{j}e_{ij}e^{-1}_{jk}=\delta_{ik}$ and $\sum_{j}e_{ij}^{-1}e_{jk}=\delta_{ik}$. This property is a straightforward generalization from finite-dimensional to infinite-dimensional linear algebra.}
		\begin{equation}
			\int_0^\infty \dd u e(\mu;u)e^{-1}(u;\mu') = \delta(\mu-\mu'),\>\>\> 
			\int_0^\infty \dd \mu e^{-1}(u;\mu)e(\mu;u') = \delta(u-u').
			\label{eq:eigenvectors_def}
		\end{equation}

		With these notations, the memory kernel and the noise statistical properties are respectively given by 
		\begin{box_normal}
			\begin{equation}
				\mathcal{M}(\mft) := \int_0^\infty \nu(\mu)e^{-\mu \mft}\dd \mu, \>\> \nu(\mu):= \int_{0}^\infty  \kappa(\mu)\Upsilon(u)e(\mu;u)\dd u, \>\>\> \kappa(\mu): = \int_0^\infty e^{-1}(u;\mu)\dd u 
				\label{ghbgqb1tqa}
			\end{equation}
			\begin{equation}
				\la \heta_{\mft}\ra = 0, \>\>\>
				\la \heta_{\mft_1}\heta_{\mft_2}\ra = \sigma^2 \left[\delta(\mft_1-\mft_2) + \mathcal{M}(|\mft_1-\mft_2|)+\int_0^\infty \dd \mft'\mathcal{M}(\mft')\mathcal{M}(\mft'+|\mft_1-\mft_2|)\right]. 
			\end{equation}
		\end{box_normal}
		
	\subsubsection*{Derivation.}
		Let us rewrite expression (\ref{eq:GaussianSPDE_SSE_trans_2}) as  
		\begin{equation}
			\frac{\partial \hZ_{\mft}(u)}{\partial \mft} = -\int_0^\infty K(u,u')\hZ_{\mft}(u')\dd u' - \gamma \hV_{\mft} + \sigma \hxi^{\mrG}_{\mft}, \>\>\> K(u,u') = u\delta (u-u') + \Upsilon(u').
		\end{equation}
		By introducing the representation based on the eigenvectors $\{e(\mu;u)\}_{\mu}$ of $K(u,u')$
		\begin{equation}
			\hw_{\mft}(\mu) := \int_0^\infty e^{-1}(u;\mu)\hZ_{\mft}(u)\dd u 
		\end{equation}
		we formally obtain the explicit representation of $\hZ_{\mft}(u)$,
		\begin{equation}
			\hZ_{\mft}(u) = \int \hw_{\mft}(\mu)e(\mu;u)\dd \mu,
			\label{eq:inverse_trans}
		\end{equation} 
		where we used Eq.~\eqref{eq:eigenvectors_def}.
		We thus obtain
		\begin{equation}
			\frac{\partial \hw_{\mft}(\mu)}{\partial \mft} = -\mu \hw_{\mft}(\mu) +\kappa(\mu)\left(-\gamma \hV_{\mft} + \sigma \hxi^{\mrG}_{\mft}\right), \>\>\> \kappa(\mu): = \int_0^\infty e^{-1}(u;\mu)\dd u,
		\end{equation}
		whose solution is given by
		\begin{equation}
			\hw_{\mft}(\mu) = \kappa(\mu)\int_{-\infty}^{\mft} e^{-\mu (\mft-\mft')}\left(\sigma \hxi^{\mrG}_{\mft'}-\gamma \hV_{\mft'}\right)\dd \mft',
		\end{equation}
		which leads to the explicit form of $\hZ_{\mft}$ as 
		\begin{equation}
			\hZ_{\mft}(u) = \int \dd \mu e(\mu;u) \kappa(\mu)\int_{-\infty}^{\mft} \dd \mft' e^{-\mu (\mft-\mft')}\left(\sigma \hxi^{\mrG}_{\mft'}-\gamma \hV_{\mft'}\right)
			\label{eq:hw_sol_trans}
		\end{equation}
		from Eq.~\eqref{eq:inverse_trans}. From Eqs.~\eqref{eq:GaussianSPDE_SSE_trans_1} and \eqref{eq:hw_sol_trans}, we obtain
		\begin{align}
			\frac{\dd\hV_{\mft}}{\dd \mft} = \sigma \hxi^{\mrG}_{\mft} - \gamma \hV_{\mft} - \int_{-\infty}^{\mft} \dd \mft' \left[\int_0^\infty \dd u\int_{0}^\infty \dd \mu \kappa(\mu)\Upsilon(u)e(\mu;u)e^{-\mu(\mft-\mft')} \right]\left(\sigma\hxi^{\mrG}_{\mft'}-\gamma \hV_{\mft'}\right).
		\end{align}
		This equation can be written as
		\begin{equation}
			\frac{\dd \hV_{\mft}}{\dd \mft} = -\gamma \hV_{\mft} -\int_{-\infty}^{\mft} \mathcal{M}(\mft-\mft')\hV_{\mft'}\dd \mft' + \heta_{\mft}
		\end{equation}
		with the memory kernel 
		\begin{equation}
			\mathcal{M}(\mft) := \int_0^\infty  \nu(\mu)e^{-\mu \mft}\dd\mu, \>\> \nu(\mu):= \int_{0}^\infty  \kappa(\mu)\Upsilon(u)e(\mu;u)\dd u
		\end{equation}
		and the colored Gaussian noise\footnote{Any noise composed of a sum of Gaussian random numbers obeys the Gaussian statistics~\cite{KuboB}.}
		\begin{equation}
			\heta_{\mft} := \sigma \hxi^{\mrG}_{\mft} + \sigma \int_{-\infty}^{\mft} \dd \mft' \mathcal{M}(\mft-\mft')\hxi_{\mft'}^{\mrG}.
		\end{equation}

\section{Application 2: price dynamics based on nonlinear Hawkes processes}\label{sec:NLHawkesPricing}
	In this section, we illustrate another application of our formalism. We focus on modelling financial price dynamics based on a nonlinear Hawkes process. Linear Hawkes processes have become popular in econophysics as well as in econometrics of market microstructure. 

	\subsection{Model}
		\begin{figure*}
			\centering
			\includegraphics[width=170mm]{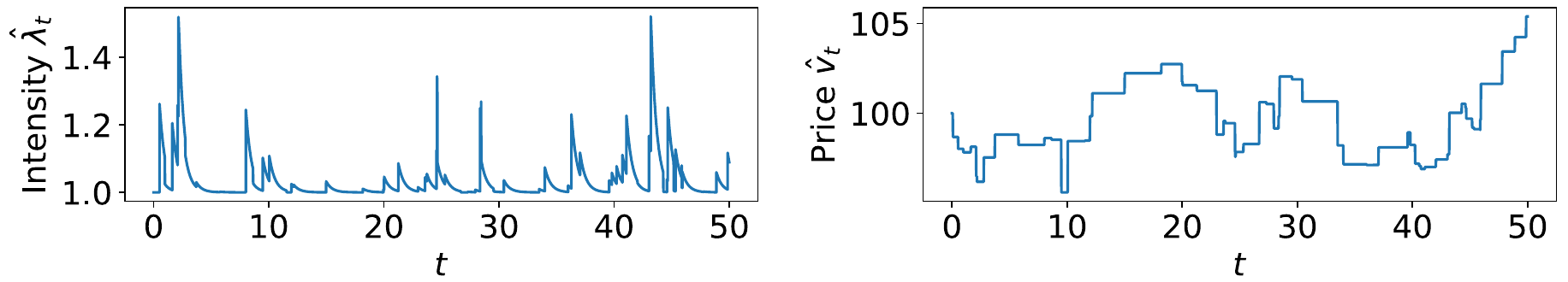}
			\caption{
				Schematic paths of the intensity $\hlambda_t$ and the price $\hv_t$ described by the nonlinear Hawkes model~\eqref{eq:NLHawkes_price} for the financial price dynamics. 
			}
			\label{fig:main:HawkesPricing}
		\end{figure*}
		Let us consider a stochastic financial model based on the nonlinear Hawkes processes, which has recently become popular to describe the price dynamics of financial assets~\cite{Blanc2017,Aubrun2023}. Let us denote $\hv_t$ the logarithm of the price of some stock at time $t$. The price dynamics is given by
		\begin{subequations}
			\label{eq:NLHawkes_price}
			\begin{equation}
				\frac{\dd \hv_t}{\dd t} = \sum_{k=1}^{\hN_t} \hy_k \delta(t-\htt_k),
			\end{equation}
			where $\hv_t$  is $k$th jump size of the log price occurring at time  $\htt_k$.
			The amplitude of the jumps are independently and identically distributed with mark distribution $\rho(y)$.
			The sequence of jumps defines the jump size series $\{\hy_k\}_k$ and the jump time series $\{\htt_k\}$. We denote by $\hN_t$ the total number of jumps during $[0,t)$. 
			We assume that both excitatory and inhibitory effects are balanced, which is realised when the mark distribution is symmetric: 
			\begin{equation}
				\rho(y) = \rho(-y).
			\end{equation}			
			The intensity $\hlambda_t$ of the jumps is assumed to obey the nonlinear Hawkes process
			\begin{equation}
				\hlambda_t = g\left(\sum_{k=1}^{\hN_t}\hy_k h(t-\htt_k)\right)
			\end{equation}
			with non-negative intensity function $g>0$ and memory kernel $h(t)$. 
		\end{subequations}
		Recall that the intensity $\hlambda_t$ gives the probability per unit time for the next jump to occur: $\hlambda_t \dd t$ is the probability 
		for the next jump to occur during $[t,t+\dd t)$. See Fig.~\ref{fig:main:HawkesPricing} for the schematic paths of this model regarding the intensity $\hlambda_t$ and the price $\hv_t$.

		This model is an example of a history-dependent Poisson processes. Indeed, the following specific history-dependent Poisson process
		\begin{equation}
			\frac{\dd \hv_t}{\dd t} = \hxiCP_{\lambda(y|\{\hv_{\tau}\}_{\tau\leq t})}, \>\>\>
			\lambda(y|\{\hv_{\tau}\}_{\tau \leq t}) := \rho(y)g\left(\int_{0}^{\infty} h(\tau)\hta_{t-\tau}\dd \tau \right), \>\>\> \hta_t := \frac{\dd \hv_t}{\dd t}
		\end{equation}
		is equivalent to the nonlinear Hawkes price model~\eqref{eq:NLHawkes_price} \cite{KzDidier2021PRL,KzDidier2023PRR}.

	\subsection{Markovian embedding}
		Our Laplace-convolution Markovian-embedding scheme~\eqref{eq:set_complete_dynamics} fully converts the nonlinear non-Markovian Hawkes process~\eqref{eq:NLHawkes_price} into a Markovian field process. Indeed, by decomposing the memory kernel as the sum of exponentials
		\begin{equation}
			h(t) := \int_0^\infty e^{-st}\tilh (s)\dd s,
		\end{equation}
		the conditional intensity can be rewritten as 
		\begin{equation}
			\lambda [y|\Gamma] = \rho(y) g\left(\int_0^\infty \tilh (s)\hz_t(s)\dd s\right). 
		\end{equation}
		This is equivalent to the Markov-embedding representation introduced in our previous works~\cite{KzDidier2019PRL,KzDidier2019PRR,KzDidier2021PRL,KzDidier2023PRR}.  

	\subsection{Field master equation}
		The field ME for the nonlinear Hawkes price model~\eqref{eq:NLHawkes_price} is
		\begin{box_normal}
			\begin{subequations}\label{eq:field_master_NLHawkes}
			\begin{equation}
				\frac{\partial P_t[\Gamma]}{\partial t} = \int_0^\infty \ds \frac{\delta}{\delta z(s)}\left(sz(s)P_t[\Gamma]\right) + 
				\int_{-\infty}^{\infty}\dy\rho(y)\left[G[z-y\bm{1}]P_t[\Gamma-\Delta \Gamma_y]-G[z]P_t[\Gamma]\right]
			\end{equation}
			with 
			\begin{equation}
				G[z]:= g\left(\int_0^\infty \tilh (s)\hz_t(s)\dd s\right) ~~{\rm and}~~\hGamma_t=(\hv_t, \{\hz_t(s)\}_{s})~.
			\end{equation}				
			\end{subequations}
		\end{box_normal}
		By integrating out both sides over $\hv_t$, this field ME reduces to the field ME for a marginal PDF $P_t[z]:=\int P_t[\Gamma]\dd x$ that was introduced in our previous works~\cite{KzDidier2019PRL,KzDidier2019PRR,KzDidier2021PRL,KzDidier2023PRR} (see Appendix~\ref{sec:app:PreviousFieldME} for the explicit derivation). In addition, the reduced field ME has been analytically solved in Refs.~\cite{KzDidier2019PRL,KzDidier2019PRR,KzDidier2021PRL,KzDidier2023PRR} for the asymptotic intensity PDF in the steady state.

	\subsection{Diffusive approximation}
		Let us apply the diffusive approximation by using the KM series \eqref{eq:KM_expansion_formula} for the 
		 the field ME~\eqref{eq:field_master_NLHawkes} and truncating it at 
		 the second order. This leads to the following approximate Fokker-Planck equation  
		\begin{equation}
			\frac{\partial P_t[\Gamma]}{\partial t} \simeq \int_0^\infty \ds \frac{\delta}{\delta z(s)}\left(sz(s)P_t[\Gamma]\right) + 
				\frac{1}{2}\left(\frac{\pd}{\pd v}+\int_0^\infty \dd s \frac{\delta}{\delta z(s)}\right)^2\alpha_2[\Gamma]P_t[\Gamma]
		\end{equation}
		with 
		\begin{equation}
			\alpha_2[\Gamma] := \sigma^2 g\left(\int_0^\infty \tilh (s)z(s)\dd s\right), \>\>\> 
			\sigma^2 := \int_{-\infty}^\infty y^2\rho(y)\dd y.
		\end{equation}
		This field Fokker-Planck equation is equivalent to 
		\begin{equation}
			\frac{\dd \hv_t}{\dd t} = D_t \hxi^{\mrG}_t, \>\>\> 
			D_t := \sigma^2 \hlambda_t~.
			\label{thyhwqgqt|}
		\end{equation}
This recovers the standard Geometric Brownman Motion model of price dynamics for constant $\hlambda_t$.
For non constant $\hlambda_t$, equation (\ref{thyhwqgqt|}) recovers the general class of stochastic volatility models \cite{Bookstovoe}.
Here, we derived that the volatility is proportional to the intensity $\hlambda_t$ of the underlying point process.
In other words, our nonlinear Hawkes (\ref{eq:NLHawkes_price}) combined with our Markovian embedding and the
diffusive approximation provide an interpretation of the source of stochastic volatility, which is here 
interpreted as resulting from the underlying jump intensity and its nonlinear memory structure.

\section{Discussion}\label{sec:discussion}
	This section delves into the ramifications of our research and outlines our perspective on several outstanding technical challenges yet to be addressed.
	
		\subsection{Comparison with the projection-operator formalism}
			Our formulation bears similarities to the projection-operator formalism, as both theories pertain to the derivations of the Generalized Langevin Equations (GLEs). In this subsection, we juxtapose the two approaches, evaluating their respective advantages and disadvantages.

			The projection-operator formalism originated in the 1950s and 1960s, crafted by pioneers like Nakajima, Mori, Zwanzig, and Kawasaki~\cite{Nakajima1958,Mori1965,Zwanzig1960,Zwanzig1961,ZwanzigTB,EvansMorrissB}. Particularly, Mori's approach focuses on establishing a microscopic foundation for the Generalized Langevin Equations (GLEs). In the projection-operator formalism, the selection of several slow variables is necessitated, guided by physical intuitions or empirical findings, as these variables cannot be determined theoretically. Subsequently, a projection operator is defined to dissect the phase-space dynamics between the function space, exclusive to slow variables, and the remainder.

			Through the application of integral identities associated with projection operators, GLEs are derived. A notable merit of this approach is the formal derivation of GLEs from microscopic dynamics, providing a rigorous connection to underlying physical processes. However, a significant drawback lies in the inherent ambiguity of the approximation involved. While all calculations are theoretically exact, eliciting nontrivial predictions mandates the approximate computation of noise statistics and friction coefficients. This level of approximation is notably more intricate compared to conventional statistical-physics theories. In fact, the determination of theoretical key perturbation/control parameters for the conclusive deduction of the GLEs from microscopic dynamics remains unambiguous, making this process elusive.

			Within the foundational framework of statistical physics pertaining to the GLEs, a drawback of our theory is the requisite assumption of the one-dimensional non-Markovian jump process~\eqref{eq:def_history_dependent_Poisson} as an initial standpoint. This assumption is fundamentally heuristic, primarily rooted in phenomenological considerations. Conversely, a significant advantage of our approach is the explicit definition of the key perturbation parameter. Specifically, the small-jump scaling parameter, $\eps$ -- generally anticipated to represent the mass ratio between the Brownian particle and surrounding entities -- serves a crucial and explicit role in our asymptotic computations. This is particularly coherent for modeling dynamics of massive Brownian particles. In this sense, we have successfully established the GLEs through a physically plausible coarse-graining process, pinpointing the essential control parameter for mathematical derivation, a contrast to the methodologies embedded in the projection-operator formalism.

		\subsection{Future issue 1: physical validation of the non-Markovian jump model}
			Our theory is premised on the non-Markovian jump model~\eqref{eq:def_history_dependent_Poisson}. While intrinsic to models in seismic activity, finance, and social science -- for instance, the Hawkes process is a subset of this model -- its applicability in physics remains indeterminate. Addressing this uncertainty will necessitate further theoretical or data-driven analysis in the future.

From a theoretical standpoint, the Markovian ME formalism~\eqref{eq:review:ME_standardForm} has been substantiated in the dynamics of Brownian particles amid dilute gases~\cite{ResiboisTB,HansenTB,Spohn1980}. Indeed, the linearized Boltzmann equation, derivable from Newtonian microscopic dynamics through the Bogoliubov-Born-Green-Kirkwood-Yvon hierarchy~\cite{ResiboisTB,HansenTB} in the low-density limit, is an instance of a Markovian jump process, thus allowing systematic theoretical validation of the Markovian ME formalism~\eqref{eq:review:ME_standardForm} via kinetic theory.

Conversely, theoretical validation for the non-Markovian jump model~\eqref{eq:def_history_dependent_Poisson} is yet to be achieved. Formulating a statistical physics theory analogous to the Markovian kinetic theory to derive the non-Markovian jump process~\eqref{eq:def_history_dependent_Poisson} from microscopic Hamiltonian dynamics is imperative.
			
		\subsection{Future issue 2:  time-reversal symmetry of our field master equation}
			\begin{figure}
				\centering
				\includegraphics[width=70mm]{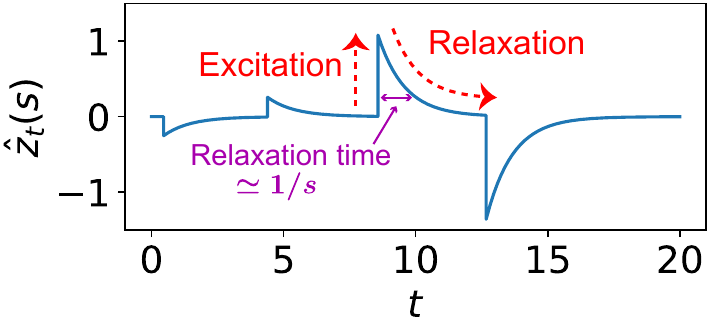}
				\caption{
					Absence of time-reversal symmetry for the auxiliary field variables $\{\hz_t(s)\}$. According to the SPDE~\eqref{eq:set_complete_dynamics_hdCP}, the path of $\hz_t(s)$ responds to the excitations due to jumps and then relaxes toward zero. The relaxation is time-irreversible, and the auxiliary field variables have thus no time-reversal symmetry. 
				}
			\end{figure}
			Exploring time-reversal symmetry is crucial when examining stochastic dynamics influenced by equilibrium fluctuations. Regrettably, this symmetry is not upheld for the field ME~\eqref{eq:field_master_hdCP}. The indispensable condition for general master equations, fully detailed in Gardiner's textbook~\cite{GardinerB} and Appendix~\ref{sec:app:detailedbalance}, is invariably breached in our field ME~\eqref{eq:field_master_hdCP}.

			The absence of time-reversal symmetry in our Laplace-type embedding representation can be intuitively understood by considering a typical path of $\{\hz_t(s)\}_{s}$. Indeed, the SPDE~\eqref{eq:set_complete_dynamics_hdCP} states that the dynamics of $\hz_t(s)$ is composed of the excitation due to the Poisson jump $\hxi^{\rm CP}_{\lambda[y|\hGamma_t]}$ and the relaxation due to the term $-s\hz_t(s)$ with the characteristic timescale $\simeq 1/s$. Since the relaxation dynamics is time irreversible, the dynamics of $\{\hz_t(s)\}_s$ has no time-reversal symmetry by construction. 

			\begin{figure*}
				\centering
				\includegraphics[width=150mm]{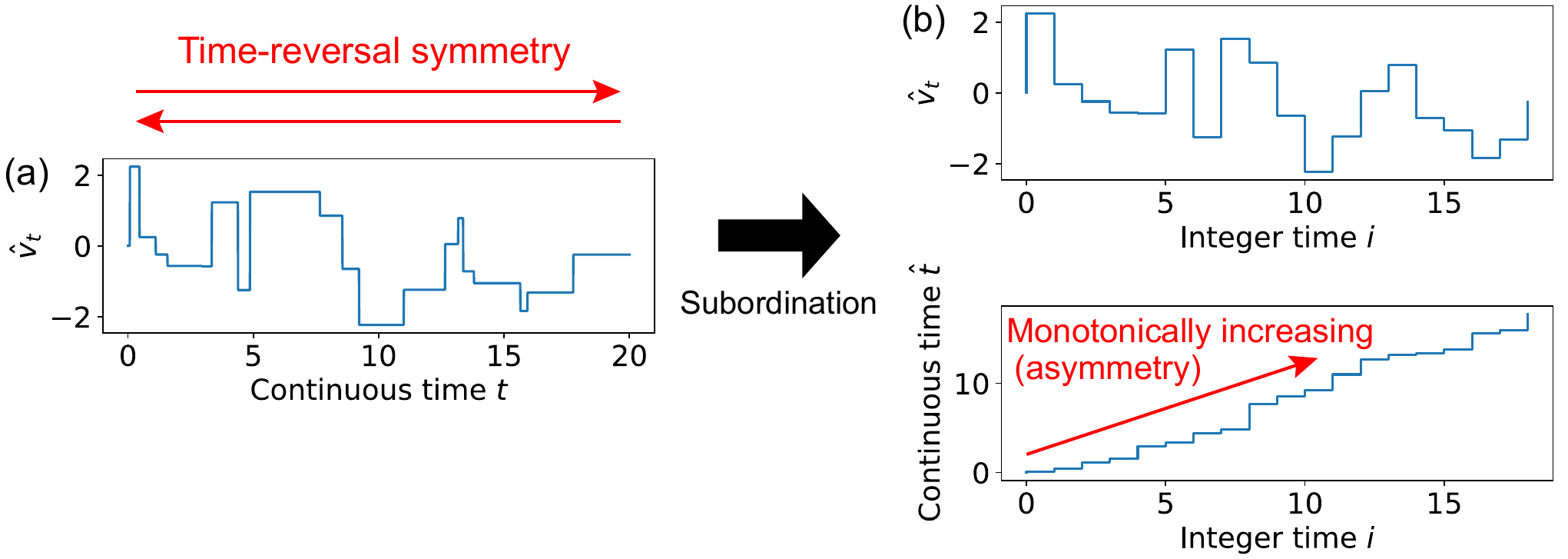}
				\caption{
					Schematic of the subordination technique. (a)~Let us consider a one-dimensional Markovian jump process with time-reversal symmetry in continuous time. (b)~We can introduce the integer time $i$, which is incremented by $1$ at each successive jump. Then, the original continuous-time one-dimensional Markovian jump process is equivalent to the discrete-time two-dimensional Markovian jump process $(\hv_i, \hat{t}_i)$, where $\hv_i$ and $\hat{t}_i$ are the velocity and the real time at the integer time $i$, respectively. By definition, $\hat{t}_i$ is monotonically increasing, and the system has no time-reversal symmetry. 
				}
				\label{fig:discussion:subordination}
			\end{figure*}
			The preservation of time-reversal symmetry in a ME is significantly contingent upon the choice of state variables~\cite{DechantKanazawaInPrep}. To illustrate, a Markovian jump model may maintain time-reversal symmetry in a one-variable representation, but it invariably loses this symmetry in a two-variable depiction. Let us assume a Markov jump process with the time-reversal symmetry (see Fig.~\ref{fig:discussion:subordination}(a)): 
			\begin{equation}
				\frac{\dd \hv_t}{\dd t} = \hxiCP_{\lambda(y|\hv_t)}, \>\>\> 
				\hxiCP_{\lambda(y|\hv_t)} = \sum_{i=1}^{\hN(t)} \hy_i\delta(t-\htt_i),
			\end{equation}
			where $\hy_i$ and $\htt_i$ are the $i$th jump size and $i$th jump time, respectively. The ME for this one-variable representation can satisfy the time-reversal symmetry by setting an appropriate intensity density $\lambda(y|v)$. 
			
			On the other hand, this process is equivalent to a discrete-time Markovian jump process with two variables 
			\begin{equation}
				\hV_{i+1} = \hV_i + \hy_i, \>\>\> \htt_{i+1} = \htt_i + \Delta \htt_i, 
			\end{equation}
			where $i$ is an integer time incremented at every jump event, $\hV_i:= \hv_{\htt_i}$ is the velocity at the $i$th jump time, and $\Delta \htt_i$ is the waiting time obeying an exponential distribution. This technique is called the {\it subordination} in the context of the continuous-time random walk theory~\cite{KlafterB} (see Fig.~\ref{fig:discussion:subordination}(b)). The ME for the two-variable representation $\hGamma_i:=(\hV_i,\htt_i)$ 
			\begin{equation}
				\Delta P_i(\hGamma) = \mathcal{L}P_i(\Gamma)
			\end{equation}
			has no time-reversal symmetry due to the monotonically increasing nature of $\htt_i$, where $\Delta P_{i}(\Gamma):=P_{i+1}(\Gamma) - P_i(\Gamma)$ is the discrete-time difference operator, and $\mathcal{L}$ is a linear operator for the discrete-time ME. 
			
			This mathematical fact suggests that there are several different Markovian-embedding formulations of the ME even if the stochastic dynamics is uniquely defined, and the time-reversal symmetry might be formulated for a specific Markovian-embedding representation. Therefore, another Markovian-embedding representation might be suitable for formulating the time-reversal symmetry. Resolving this issue will be instrumental in shaping the formulation of stochastic thermodynamics and energetics~\cite{ShiraishiB2023,KenSekimotoB2021} for non-Markovian jump processes. This is reserved for the future.

		\subsection{Future issue 3: the fluctuation-dissipation relation of the second kind}
			When the environment is in thermal equilibrium, the thermal fluctuation of the GLE must satisfies the fluctuation-dissipation relation (FDR) of the second kind: 
			\begin{equation}
				\la \heta_{\mft_1}\heta_{\mft_2}\ra = 2T\left\{\gamma\delta(\mft_1-\mft_2)+\mathcal{M}(|\mft_1-\mft_2|)\right\}
			\end{equation}
			where the left-hand side is the cross-correlation between $\heta_{\mft_1}$ and $\heta_{\mft_2}$, $T$ is the temperature 
			and we have taken units where the Boltzmann constant is unity. This fluctuation-dissipation relation of the second kind is equivalent to the time-reversal symmetry of the GLE. This was one of the most important issues for the statistical-physics foundation of the GLE particularly within the context of linear response theory~\cite{KuboB} and the projection-operator formalism~\cite{ZwanzigTB}.
			
			Since our Markovian-embedding formulation does not yet convert the time-reversal symmetry of the field ME, the necessary and sufficient condition for this fluctuation-dissipation relation of the second kind is not yet identified. The identification of these conditions is also an important future challenge. 

		\subsection{Future issue 4: formal relations to quantum field theory}
			Our field master equation (Fokker-Planck) is formally related to quantum field theory. Indeed, the field Fokker-Planck equation for the GLE with time-reversal symmetry is equivalent to a non-Hermitian quantum field theory with the Hermitian part of its Hamiltonian describing a field of harmonic oscillators~\cite{KzDidier2019PRR}. Indeed, a similar renormalisation issue appears regarding the infinite zero-point energy of the field harmonic oscillators. Since the first-order contribution of the system-size expansion for the non-Markovian jump process leads to the GLE, its next-order perturbation theory might require the use of methods developed in quantum field theory, such as the Feynman-diagram expansion. Establishing such field-theoretical techniques will be an interesting future topic.

\section{Conclusion}\label{sec:conclusion}
	In conclusion, we have introduced a comprehensive stochastic framework through a field master equation, encompassing all one-dimensional non-Markovian jump processes. Utilizing the Laplace-convolution embedding representation, we have demonstrated the transformation of any non-Markovian jump process into Markovian-field dynamics. We subsequently derived the corresponding field master equation and procured an asymptotic solution using a generalized system-size expansion. In essence, this framework can be applied to any jump processes, assuming one-dimensional dynamics are driven by collisions. We posit that this model's flexibility makes it adept at accommodating a wide array of point-process data, proving invaluable for data analyses.

\begin{acknowledgements}
	KK was supported by JST PRESTO (Grant No.~JPMJPR20M2) and JSPS KAKENHI (Grant Nos.~21H01560, 22H01141, and 23H00467). DS was partially supported by the National Natural Science Foundation of China (Grant no. U2039202), Shenzhen Science and Technology Innovation Commission (Grant no. GJHZ20210705141805017), as well as by the Feature Innovation Project of Colleges and Universities in Guangdong Province (2020WTSCX082), Shenzhen Science and Technology Innovation Commission Project (grant no. GJHZ20210705141805017 and grant no. K23405006), and the Center for Computational Science and Engineering at Southern University of Science and Technology.

	We thank Masato Itami for his careful reading of our manuscripts with useful feedback. We also thank Andreas Dechant for his crucial comment that the condition of  time-reversal symmetry explicitly depends on the selection of Markovian-embedding representations. 
	In addition, KK had a fruitful discussion with Alexander Teretenkov in the international conference Statphys 28 regarding Markovian embedding techniques in quantum stochastic processes. Furthermore, the discussions with Hiroyasu Tajima and Sosuke Ito were quite inspiring for the future application to stochastic thermodynamics. 
\end{acknowledgements}

\appendix
\section{Dirac's $\delta$ function and functional derivative}\label{sec:app:dirac-functionalderivative}
	In this Appendix, we formally define the Dirac $\delta$ function and the functional derivatives. While our formulation is systematic enough at the theoretical-physics level, presenting mathematically rigorous formulations is out of scope in this report. 
	
	\subsection{Formal definition}
		\subsubsection{Dirac's $\delta$ function}
			Dirac's $\delta$ function is formally defined by 
			\begin{equation}
				\delta(s-s') = \begin{cases}
					0 & (s\neq s') \\
					\infty & (s=s')
				\end{cases}, \>\>\> 
				\int_{-\infty}^\infty \ds f(s)\delta(s-s') = f(s')
			\end{equation}
			with any real numbers $s$ and $s'$. The $\delta$ function is the continuous analogue of the Kronecker $\delta$
			for discrete variables, which is defined by 
			\begin{equation}
				\delta_{ij}= \begin{cases}
					0 & (i\neq j) \\
					1 & (i=j)
				\end{cases}, \>\>\> 
				\sum_{k}\delta_{ik}f_k=f_i
			\end{equation}
			for any integers $i$ and $j$. 

			Dirac's $\delta$ function can be formally constructed via a lattice model. Let us discretize the real number line $(0,\infty)$, such that $s_k = k \ds$ with the lattice constant $\ds>0$ and any integer $k>0$. The Dirac $\delta$ function is formally defined by 
			\begin{equation}
				\delta(s-s') := \lim_{\ds\searrow 0} \frac{1}{\ds}\delta_{ij},
			\end{equation}
			where $s\in [s_i,s_{i+1})$ and $s'\in [s_j,s_{j+1})$. Indeed, with this definition, we obtain the consistent relationship 
			\begin{equation}
				\int_{-\infty}^\infty \ds f(s)\delta(s-s') = \lim_{\ds\searrow 0} \sum_{i} \ds_if(s_i)\frac{1}{\ds}\delta_{ij} = f(s'). 
			\end{equation}

		\subsubsection{Functional derivative}
			Let us define the functional derivatives as a formal limit from the finite-dimensional vector function (i.e., a lattice model). Let us consider the $K$-dimensional vector $\bm{z}:=(z_1,\dots, z_K)$ and an arbitrary function $f(\bm{z})$. The partial derivative of $f(\bm{z})$ is written as $(\partial f(\bm{z}))/(\partial z_k)$ for an integer $k$. 
			
			We then consider a formal continuous limit from such a finite-dimensional models. Let us introduce $s_k:=k (\ds)$ with the lattice constant $\ds>0$ for integer $k>0$, and take the continuous limit $\ds\searrow 0$ and $K\to \infty$. The functional derivative is defined by 
			\begin{equation}
				\frac{\delta f[z]}{\delta z(s)} := \lim_{\substack{\ds\searrow 0 \\ K\to \infty}}\frac{1}{\ds}\frac{\partial f(\bm{z})}{\partial z(s_k)}
				\label{def:app:functional_der}
			\end{equation}
			where $s_k=k \ds$ and $s\in [s_{k},s_{k+1})$ with an integer $k$. 

	\subsection{Useful identities}
		\subsubsection{First-order functional Taylor expansion}
			For a finite-dimensional vector function $f(\bm{z})$, the first-order Taylor expansion is given by 
			\begin{equation}
				\dd f(\bm{z}) = \sum_{k=1}^K \frac{\partial f(\bm{z})}{\partial z_k}\dd z_k,
			\end{equation}
			with $\dd f(\bm{z}) := f(\bm{z}+\dd \bm{z})-f(\bm{z})$ for infinitesimal $\dd \bm{z}:=(\dd z_1,\dots,\dd z_K)$. In the continuous limit, we apply the replacement 
			\begin{equation}
				\sum_{k=1}^K \ds[\dots] \to \int_{0}^\infty \ds[\dots], \>\>\>  
				\frac{\partial f(\bm{z})}{\partial z(s_k)} \to \ds \frac{\delta f[z]}{\delta z(s)}, \>\>\>
				\dd z_k \to \delta z(s)
				\label{eq:app:replacement}
			\end{equation}
			to obtain the first-order functional Taylor expansion
			\begin{equation}
				\delta f[z] = \int \ds \frac{\delta f[z]}{\delta z(s)}\delta z(s) + O(\delta z^2)
			\end{equation}
			with $\delta f[z]:=f[z+\delta z] - f[z]$ with infinitesimal $\delta z$.
		
		\subsubsection{Full-order functional Taylor expansion}
			For a finite-dimensional vector function $f(\bm{z})$, the full-order Taylor expansion is given by 
			\begin{equation}
				f (\bm{z}+\Delta \bm{z}) -f(\bm{z}) = \sum_{n=1}^\infty \frac{1}{n!}\left(\sum_{k=1}^K\Delta z_k\frac{\partial }{\partial z_k}\right)^n f(\bm{z})
			\end{equation}
			with $\Delta \bm{z}:=(\Delta z_1,\dots, \Delta z_K)$.
			In the continuous limit based on the formal replacement~\eqref{eq:app:replacement}, we obtain the full-order functional Taylor expansion 
			\begin{equation}
				f[z+\Delta z]-f[z] = \sum_{n=1}^\infty \frac{1}{n!}\left(\int \ds\Delta z(s)\frac{\delta }{\delta z(s)}\right)^n f[z].
			\end{equation}
			We note that this calculation can be readily generalised for a two-argument functional $f[v;z]$ with a real value $v$ and a function $\{z(s)\}_s$, such that 
			\begin{equation}
				 f[v+\Delta v
				 ;z+\Delta z] -f[v;z]= \sum_{n=1}^\infty \frac{1}{n!}\left(\Delta v \frac{\partial}{\partial v}+\int \ds\Delta z(s)\frac{\delta }{\delta z(s)}\right)^n f[v;z]
				 \label{eq:app:fullTaylorfunctional_2var}
			\end{equation}
			with small $\Delta v$ and $\{\Delta z(s)\}_s$. Particularly, the Maclaurin series is given by 
			\begin{equation}
				f[v;z] = \sum_{n=0}^\infty \frac{1}{n!}\left(v \frac{\partial}{\partial \chi}+\int \ds z(s)\frac{\delta }{\delta \zeta (s)}\right)^n f[\chi;\zeta] \bigg|_{(\chi,\{\zeta(s)\}_s)=\bm{0}},
				\label{eq:app:Mclaurin_functional_2var}
		 \end{equation}
		 where the dummy argument variables $\chi$ and $\zeta$ are introduced to distinguish the arguments involved in the derivatives from the 
		 arguments $v$ and $z$ of the function $f[v;z]$. 

		\subsubsection{Variable transformation formula}
			Let us consider a simple variable transformation
			\begin{equation}
				\tilde{s} = as
			\end{equation}
			with a positive constant $a$. Considering the definition~\eqref{def:app:functional_der}, we obtain 
			\begin{equation}
				\frac{\delta}{\delta z(s)} := \lim_{\substack{\ds \searrow 0 \\ K\to \infty}}\frac{a}{(a \ds)}\frac{\partial }{\partial z(s_k)} = a \lim_{\substack{\dd\tilde{s}\searrow 0 \\ K\to \infty}}\frac{1}{\dd \tilde{s}}\frac{\partial }{\partial z(\tilde{s}_k)} = a\frac{\delta }{\delta z(\tilde{s})},
				\label{eq:app:var_trans_functional_der}
			\end{equation}
			which leads to the invariant integral relationship, 
			\begin{equation}
				\int \ds \frac{\delta }{\delta z(s)} = \int \dd\tl{s} \frac{\delta }{\delta z(\tl{s})}.
			\end{equation}

		\subsubsection{Partial integration}
			For a finite-dimensional vector $\bm{z}$, the partial integration is given by 
			\begin{equation}
				\int_{-\infty}^\infty P(\bm{z})\frac{\partial f(\bm{z})}{\partial z_k}\dd\bm{z} 
				= -\int_{-\infty}^\infty f(\bm{z})\frac{\partial P(\bm{z})}{\partial z_k}\dd\bm{z}
			\end{equation}
			by assuming vanishing boundary conditions $\lim_{|\bm{z}|\to \infty}P(\bm{z})=0$. As a straightforward generalisation, by considering the formal definition~\eqref{def:app:functional_der}, the partial integration of a functional $f[z]$ is given by 
			\begin{equation}
				\int P[z]\frac{\partial f[z]}{\partial z(s)}\mcD z 
				= -\int f[z]\frac{\delta P[z]}{\delta z(s)}\mcD z 
				\label{eq:app:partialIntegral_functional}
			\end{equation}
			by also assuming vanishing boundary conditions.

\section{Brief review of the white Gaussian and Poisson noises}\label{app:review_whitenoise}
	\subsection{White Gaussian noise}
		Let us consider the following stochastic difference equation (SDE) with finite timestep $dt$: 
		\begin{equation}
			\hW_{t+dt} = \hW_t + \sqrt{\dd t}\heta^{\mrG}_t
		\end{equation}
		with the standard normal random variable $\heta^{\mrG}_t$ that is independent and identically distributed (IID): $\la\heta^{\mrG}_t\heta^{\mrG}_{t'} \ra = 1$ for $t=t'$ and $\la\heta^{\mrG}_t\heta^{\mrG}_{t'} \ra = 0$ for $t\neq t'$. 

		We then consider the stochastic dynamics for the infinitesimal time step limit $\dd t\to 0$ to define the {\it Wiener process} $\hW_t$. The formal derivative of the Wiener process is called the {\it white Gaussian noise}: 
		\begin{equation}
			\hxi^{\mrG}_t := \frac{\dd \hW_t}{\dd t},
		\end{equation}
		which satisfies the relationship of the white noise
		\begin{equation}
			\la \hxi^{\mrG}_t\hxi^{\mrG}_{t'}\ra = \delta(t-t').
		\end{equation}

	\subsection{White Poisson noise}
		The white Poisson noise is composed of the sum of $\delta$ functions, such that 
		\begin{equation}
			\hxi^{\mrP}_{t,\lambda(y)} := \sum_{i=1}^{\hN(t)}\hy_i \delta(t-\htt_i),
		\end{equation}
		which is characterised by the intensity density function $\lambda(y)$. $\{\htt_i\}_i$ is the time sequence of jump events, $\{\hy_i\}_i$ is the sequence of jump sizes (called {\it mark} in the context of point processes), and $\hN(t)$ is the total number of jump events during the interval $[0,t)$. The probability that an event with jump size $\hy_i \in [y,t+\dd y)$ occurs during $[t,t+\dd t)$ is given by 
		\begin{equation}
			\lambda(y)\dd y\dd t. 
		\end{equation}
		When the total intensity $\lambda_{\rm tot}:= \int_{-\infty}^\infty \lambda(y)\dd y$ is finite ($\lambda_{\rm tot}<\infty$), an event occurs during $[t,t+\dd t)$ with the probability 
		\begin{equation}
			\lambda_{\rm tot}\dd t
		\end{equation}
		and the jump size distribution is given by 
		\begin{equation}
			\rho(y) = \frac{\lambda(y)}{\int_{-\infty}^{\infty}\lambda(y)\dd y}.
		\end{equation}

	\subsection{White noise}
		The white noise $\hxi^{\rm W}(t)$ is the time-homogeneous noise without time correlation and is defined as the formal time-derivative of the L\'evy process. The L\'evy process $\hL_t$ is defined as the stochastic process satisfying the following properties: (i)~$L_0=0$. (ii)~For any $0\leq t_1< t_2 <\dots <t_n$, $L_{t_2}-L_{t_1}$, $L_{t_3}-L_{t-2}$, \dots, $L_{t_n}-L_{t_{n-1}}$ are independent of each other. (iii)~For any $s<t$, the PDF of $L_t-L_s$ is equal to that of $L_{t-s}$. With mean $m:=\la\hxi^{\rm W}\ra$, the white noise has no correlation, such that 
		\begin{equation}
			\left< \left(\hxi^{\rm W}_t-m\right)\left(\hxi^{\rm W}_{t'}-m\right) \right> = \delta(t-t').
		\end{equation}
		
		According to the L\'evy-It\^o decomposition, any white noise is decomposed of the sum of the constant drift $m$, the white Gaussian noise, and the white Poisson noise as given by Eq.~\eqref{eq:Levy-Ito_decomposition}. Thus, the white Gaussian and Poisson noises are the fundamental components of the Markovian noise sources. 

\section{Review of the system-size expansion for the Markovian jump process}
\label{sec:app:review_SSE}
	Let us briely explain the system-size expansion for the Markovian jump process~\eqref{eq:review:Markovian_Jump_for_SSE}. With the scaling assumption~\eqref{eq:review_SSE_Markov_scaling}, the master equation~\eqref{eq:review:ME_standardForm} can be rewritten as 
	\begin{equation}
		\frac{\pd P_t(v)}{\pd t} = \frac{1}{\ve}\int_{-\infty}^\infty \dd y\left[W\left(\frac{y}{\ve}\Big| v-y\right)P_t(v-y)-W\left(\frac{y}{\ve}\Big| v\right)P_t(v)\right]
		= \sum_{n=1}^\infty \frac{(-\ve)^n}{n!}\frac{\partial^n}{\partial v^n}[\mcA_n(v)P_t(v)]
		\label{eq:app:KM_SSE}
	\end{equation}
	with the transformation $y=\ve Y$ and the $\ve$-independent KM coefficient defined by 
	\begin{equation}
		\mcA_n(v) := \int_{-\infty}^\infty Y^n W(Y|v)\dd Y.
	\end{equation}
	We assume the following stability conditions around $v= 0$: 
	\begin{enumerate}
		\item \tb{Linear stability:} The first-order KM coefficient has a single stable point, such that
		\begin{equation}
			\mcA_1(0) = 0, \>\>\> \gamma := -\frac{\partial}{\partial v}\mcA_1(v)\big|_{v=0} = -\mcA_1^{(1)}(0) > 0.
		\end{equation} 
		with $\mcA_n^{(k)}(v):= \partial^k \mcA_n(v)/\partial v^k$. 
		\item \tb{Existence of the noise term:} The Gaussian noise term is assumed to be present even for $\ve\to 0$, such that 
		\begin{equation}
			\sigma^2 := \mcA_2(0) > 0.
		\end{equation}
		\item \tb{Scaled variables:} Furthermore, we apply the transformation of variables: 
		\begin{equation}
			\mft := \ve t, \>\>\> V := \frac{v}{\sqrt{\ve}}.
		\end{equation}
		These scaled variables are introduced to focus on the long-time limit (i.e., $\mft=O(1)\Longleftrightarrow t=O(\ve^{-1})\gg 1$) and to enlarge the peak of the velocity PDF (i.e., $V=O(1)\Longleftrightarrow v=O(\ve^{1/2})\ll 1$) in the small-noise limit. 
	\end{enumerate}
	With these assumptions, the KM series~\eqref{eq:app:KM_SSE} can be rewritten as 
	\begin{align}
		\frac{\pd P_t(V)}{\pd \mft} &= 
		\sum_{n=1}^\infty \sum_{k=0}^\infty\frac{\partial^n}{\partial V^n} \left[\sum_{k=1}^\infty V^k\frac{(-1)^n\ve^{\frac{k+n}{2}-1}}{n!}\frac{\mcA^{(k)}_n(0)}{k!}P_t(V)\right] \notag \\
		&= \gamma \frac{\pd}{\pd V}[VP_t(V)] + \frac{\sigma^2}{2}\frac{\pd^2}{\pd V^2}P_t(V) + o(\ve^{1/2})
	\end{align}
	where we applied the Taylor expansion of the $n$th-order KM coefficient
	\begin{equation}
		\mcA_n(v) = \sum_{k=0}^\infty \frac{\ve^{k/2}V^{k}}{k!}\mcA^{(k)}_n(0). 
	\end{equation}
	In the small-noise limit $\ve \to 0$, we obtain the FP equation 
	\begin{equation}
		\frac{\pd P_t(V)}{\pd \mft} = \gamma \frac{\pd}{\pd V}[VP_t(V)] + \frac{\sigma^2}{2}\frac{\pd^2}{\pd V^2}P_t(V),
	\end{equation}
	which is equivalent to the Langevin equation 
	\begin{equation}
		\frac{\dd \hV}{\dd \mft} = -\gamma \hV + \sigma \hxi^{\rG}.
	\end{equation}

\section{Trivial Markovian embedding for discrete-time stochastic processes}\label{sec:app:DiscreteTime_Embedding}
	Here we show a trivial approach of Markovian embedding available only for discrete-time stochastic processes. 
	
	\subsection{Discrete-time stochastic process and Markovian embedding}
		Let us consider a discrete-time stochastic difference equation, 
		\begin{equation}
			\hx_{t+1} = f(\hx_{t},\hx_{t-1},\dots ,\hx_{t-K})
			\label{eq:app:naive_embedding_discrete}
		\end{equation}
		with a positive integer $K>0$. We assume $f$ includes noise terms in general and can be stochastic, such as the ARIMA model. 
		
		This model can be trivially converted onto Markovian dynamics by introducing the phase-space vector
		\begin{equation}
			\hat{\bm{\Gamma}}_t:= (\hx_{t},\hx_{t-1},\dots, \hx_{t-K})^{\rm T}.
		\end{equation}
		with the superscript ${\rm T}$ signifies the transpose operator. Indeed, we obtain a first-order stochastic difference equation
		\begin{equation}
			\hat{\bm{\Gamma}}_{t+1} = S \hat{\bm{\Gamma}}_t + \bm{f}(\hat{\bm{\Gamma}}_t), \>\>\> 
			S := 
			\begin{pmatrix}
				0 & 0 & \dots & 0 & 0\\
				1 & 0 & \dots & 0 & 0\\
				0 & 1 & \dots & 0 & 0\\
				\vdots & \vdots & \ddots & \vdots & \vdots\\
				0 & 0 & \dots & 1 & 0
			\end{pmatrix}, \>\>\> 
			\bm{f}(\hat{\bm{\Gamma}}_t) := 
			\begin{pmatrix}
				f(\hGamma_t) \\ 
				0 \\ 
				0 \\
				\vdots \\ 
				0
			\end{pmatrix}.
			\label{eq:app:embedding_naive}
		\end{equation}
		Here $S$ is the finite-dimensional shifting operator, such that 
		\begin{equation}
			\begin{pmatrix}
				0 \\ 
				\hx_{t} \\ 
				\hx_{t-1} \\
				\vdots \\ 
				\hx_{t-K-1}
			\end{pmatrix}
			= 
			\begin{pmatrix}
				0 & 0 & \dots & 0 & 0\\
				1 & 0 & \dots & 0 & 0\\
				0 & 1 & \dots & 0 & 0\\
				\vdots & \vdots & \ddots & \vdots & \vdots\\
				0 & 0 & \dots & 1 & 0
			\end{pmatrix}
			\begin{pmatrix}
				\hx_{t} \\ 
				\hx_{t-1} \\ 
				\hx_{t-2} \\
				\vdots \\ 
				\hx_{t-K}
			\end{pmatrix}.
		\end{equation}
		Similar ideas are used in Econometrics~\cite{JDHamilton} regarding the lag operator. If the time is discrete, this formulation can be straigtforwardly generalised even for $K\to \infty$, where the embedding dimension is infinite and thus the dynamics is truly non-Markovian. 

	\subsection{Technical contribution of the Laplace-convolution representation}
		This fact implies that Markovian embedding is trivial for discrete-time stochastic processes. However, a straightforward generalisation of this specific embedding is difficult for continuous-time stochastic processes. Indeed, it is challenging to generalise the shifting operator $S$ for continuous-time representations, even at a formal level. 

		Let us attempt to write the formal continuous representation from the naive discrete-time embedding equation~\eqref{eq:app:embedding_naive}. By considering the continuous limit with $K\to\infty$ and $\dd t\to 0$ for the time interval, let us write the phase-space vector as $\{\hGamma_t(s)\}_{s\geq 0}$ parametrised with $s\geq 0$ defined by 
		\begin{equation}
			\hGamma_t(s) := \hx_{t-s}.
		\end{equation}
		Equation~\eqref{eq:app:embedding_naive} can be formally written as 
		\begin{equation}
			\hGamma_{t+\dd t}(s) = \int_{0}^\infty \dd s'\delta(s-\dd t-s') \hGamma_{t}(s') + f[\hGamma_t]\delta_{s,0},
		\end{equation}
		or equivalently, 
		\begin{equation}
			\frac{\dd \hGamma_{t}(s)}{\dd t} = \int_{0}^\infty \dd s' K(s,s') \hGamma_t(s') + f[\hGamma_t]\delta_{s,0}, \>\>\> 
			K(s,s') := \frac{\dd}{\dd s'}\delta(s-s').
		\end{equation}
		This equation does not make sense even at the theoretical physics level due to the apparent singularity of the $ \delta $ function and its derivative. Thus, the naive embedding~\eqref{eq:app:naive_embedding_discrete} for the discrete-time processes cannot be straightforwardly generalised toward the continuous-time processes, even at the formal level.

		The Laplace-convolution representation technically solves this problem. The shifting operator $S$ has an analytically tractable representation in the Laplace-convolution space, and, thus, the original non-Markovian dynamics is mapped onto a first-order Markovian SPDE.

\section{Eigenvalues and eigenfunctions of the matrix $K(u,u')$}\label{sec:app:eigenvalues}
	Let us prove that all the eigenvalues of $K(u,u')$ defined by  Eqs.~\eqref{def:K_matrix} are real and positive. We define 
	\begin{equation}
		\mcK(u,u'):= \sqrt{\frac{\Upsilon(u)}{\Upsilon(u')}}K(u,u') = \sqrt{uu'}\delta (u-u') + \sqrt{\Upsilon(u)\Upsilon(u')},
		\label{def:app:K_tilde}
	\end{equation}
	where $\Upsilon(u)$ is defined by Eqs.~\eqref{def:gamma}. Since $\mcK$ is symmetric ($\mcK(u,u')=\mcK(u',u)$), its eigenvalues $\tl{\mu}$ are real, such that 
	\begin{equation}
		\int_0^\infty \dd u'\mcK(u,u')\tl{e}(\tl{\mu};u') = \tl{\mu} \tl{e}(\tl{\mu};u), \>\>\> \tl{\mu} \in \bm{R}
	\end{equation}
	with the eigenfunctions $\{\tl{e}(\tl{\mu};u)\}_{\tl{\mu}}$. In addition, we find a positive-definite inequality for any function $f(u)$, such that 
	\begin{equation}
		\int_0^\infty f(u_1)\mcK(u_1,u_2)f(u_2) \dd u_1 \dd u_2 = \int_{0}^\infty u f^2(u) \dd u  + \left(\int_0^\infty \sqrt{\Upsilon(u)}f(u)\right)^2 \dd u > 0,
	\end{equation}
	except for the trivial case $f(u)=0$ for all $u$. This implies that the symmetric ``matrix" $\mcK(u,u')$ is positive definite, and thus has only real eigenvalues. In addition, since all the eigenvalues are positive for the symmetric ``matrix" $\mcK(u,u')$, it has an inverse matrix. 
	
	Finally, from the definition~\eqref{def:app:K_tilde}, we find that  
	\begin{equation}
		\int_0^\infty \dd u'K(u,u')\frac{\tl{e}(\tl{\mu};u')}{\sqrt{\Upsilon(u')}} = \tl{\mu} \frac{\tl{e}(\tl{\mu};u)}{\sqrt{\Upsilon(u)}},
	\end{equation} 
	implying that all the eigenvalues of $K(u,u')$ correspond to those of $\mcK(u,u')$, such that
	\begin{equation}
		\mu = \tl{\mu}, \>\>\> e(\mu;u) = \frac{\tl{e}(\tl{\mu};u)}{\sqrt{\Upsilon(u)}}.
	\end{equation}
	This means that all the eigenvalues of $K(u,u')$ are real and positive. Furthermore, $K(u,u')$ has an inverse matrix.

\section{Explicit relation with the field master equation for nonlinear Hawkes processes previously derived in Refs.~\cite{KzDidier2021PRL,KzDidier2023PRR}}\label{sec:app:PreviousFieldME}
	The field ME~\eqref{eq:field_master_NLHawkes} can be easily transformed. Let us define the following quantities: 
	\begin{equation}
		s:= \frac{1}{x}, \>\>\> 
		z'(x):= \tilde{h}'(x) z(s), \>\>\> 
		\tilde{h}'(x):= \frac{\tilde{h}(s)}{x^2} ,
	\end{equation}
	satisfying $h(t)=\int_0^\infty \tilde{h}'(x)e^{-t/x}\dd x$ and $G'[z']=g\left(\int_0^\infty \tilde{z}'(x)\dd x\right)$. By integrating $P_t[\Gamma]$ over $\hv$ to define the marginal PDF
	\begin{equation}
		P_t[z']:= \int P_t[v;z]\dd v,
	\end{equation}
	we obtain from equation~\eqref{eq:field_master_NLHawkes}
	\begin{equation}
		\frac{\partial P_t[z']}{\partial t} = \int \dd x \frac{\delta}{\delta z'(s)}\left(\frac{z'(s)}{x}P_t[z']\right) + \int_{-\infty}^\infty \dd y \rho(y)G'[z'-y\tilde{h}']P_t[z'-y\tilde{h}']-G'[z']P_t[z'].
	\end{equation}
	This equation is equivalent to the field ME in Refs.~\cite{KzDidier2021PRL,KzDidier2023PRR}.

	\section{Time-reversal symmetry of the master equation}\label{sec:app:detailedbalance}
		We review the necessary and sufficient condition of the validity of time-reversal symmetry according to Ref.~\cite{GardinerB}. For a finite-dimensional Markovian stochastic processe $\bm{x}:=(x_1,\dots,x_K)^{\rm T}$, the general master equation is given by 
		\begin{equation}
			\frac{\partial P_t(\bm{x})}{\partial t} = \left[-\sum_{k} \frac{\partial}{\partial x_k}A_k(\bm{x})+\frac{1}{2}\sum_{k,l}\frac{\partial^2}{\partial x_k \partial_l}B_{kl}(\bm{x})\right]P_t(\bm{x}) + \int \dd \bm{y}\left[\lambda(\bm{x}|\bm{y})P_t(\bm{y})-\lambda(\bm{y}|\bm{x})P_t(\bm{x})\right],
		\end{equation}
		where $A_k$ is the drift term, $B_{kl}\geq 0$ is the diffusion term, and $\lambda(\bm{y}|\bm{x})\geq 0$ is the jump-intensity density for jumps from $\bm{x}$ to $\bm{y}$. 
	
		Let us define the time-reversal operator $\eps_k$ such that 
		$\eps_k = 1$ if $x_k$ is an even variable, and $\eps_k = -1$ if $x_k$ is an odd variable. Typically, the velocity (position) is an odd (even) variable because it has the odd (even) parity under time reversal. The necessary and sufficient condition for time-reversal symmetry to hold is given by 
		\begin{subequations}
		\begin{align}
			\lambda(\bm{y}|\bm{x})P_{\rm ss}(\bm{x}) &= \lambda(\bm{\eps}\bm{x} | \bm{\eps}\bm{y})P_{\rm ss}(\bm{y}) \\
			\eps_k A_k (\bm{\eps}\bm{x}) P_{\rm ss}(\bm{x}) &= -A_k(\bm{x})P_{\rm ss}(\bm{x}) + \sum_l \frac{\partial}{\partial x_l}[B_{kl}P_{\rm ss}(\bm{x})] \\ 
			\eps_k\eps_l B_{kl}(\bm{\eps}\bm{x}) &= B_{kl}(\bm{x})
		\end{align}
		with 
		\begin{equation}
			\bm{\eps} := \begin{pmatrix}
				\eps_1 & 0 & \dots & 0 \\
				0 & \eps_2 & \dots & 0 \\
				\vdots & \vdots & \ddots & \vdots \\
				0 & 0 & \dots & \eps_K 
			\end{pmatrix}.
		\end{equation}
		\end{subequations}


\end{document}